\documentclass[journal]{IEEEtran}
\usepackage[utf8]{inputenc}
\usepackage{amsmath,amssymb,amsfonts}
\usepackage{mathtools}
\usepackage{bbm}
\usepackage{cite}
\usepackage{graphicx}
\usepackage{subcaption}
\usepackage{booktabs}
\usepackage{tabularx}
\usepackage{multirow}
\usepackage{array}
\newcolumntype{L}[1]{>{\raggedright\arraybackslash}m{#1}}

\usepackage{enumitem}

\usepackage{algorithm}
\usepackage{algpseudocode}

\usepackage{xcolor}
\usepackage{xurl}
\usepackage{microtype}

\usepackage{tikz}
\usetikzlibrary{decorations.pathreplacing}
\usetikzlibrary{positioning,arrows.meta,quotes}
\usetikzlibrary{shapes,snakes}
\usetikzlibrary{bayesnet}
\tikzset{>=latex}
\tikzstyle{plate caption} = [caption, node distance=0, inner sep=0pt, below left=5pt and 0pt of #1.south]

\usepackage{cite}

\usepackage[hidelinks]{hyperref}

\title{Active Simulation-Based Inference for Scalable Car-Following Model Calibration}

\author{Menglin~Kong, 
        Chengyuan~Zhang, 
        and~Lijun~Sun%
\thanks{M. Kong, C. Zhang, and L. Sun are with the Department of Civil Engineering,
McGill University, Montreal, QC H3A 0C3, Canada.}%
\thanks{Corresponding author: Chengyuan Zhang (e-mail: enzozcy@gmail.com).}
}

\begin{document}
\maketitle

\begin{abstract}
Credible microscopic traffic simulation requires car-following models that capture both the average response and the substantial variability observed across drivers and situations. However, most data-driven calibrations remain deterministic, producing a single best-fit parameter vector and offering limited guidance for uncertainty-aware prediction, risk-sensitive evaluation, and population-level simulation. Bayesian calibration addresses this gap by inferring a posterior distribution over parameters, but per-trajectory sampling methods such as Markov chain Monte Carlo (MCMC) are computationally infeasible for modern large-scale naturalistic driving datasets.
This paper proposes an active simulation-based inference framework for scalable car-following model calibration. The approach combines (i) a residual-augmented car-following simulator with two alternatives for the residual process and (ii) an amortized conditional density estimator that maps an observed leader--follower trajectory directly to a driver-specific posterior over model parameters with a single forward pass at test time. To reduce simulation cost during training, we introduce a joint active design strategy that selects informative parameter proposals together with representative driving contexts, focusing simulations where the current inference model is most uncertain while maintaining realism.
Experiments on the HighD dataset show improved predictive accuracy and closer agreement between simulated and observed trajectory distributions relative to Bayesian calibration baselines, with convergence and ablation studies supporting the robustness of the proposed design choices. The framework enables scalable, uncertainty-aware driver population modeling for traffic flow simulation and risk-sensitive transportation analysis.
\end{abstract}


\section{Introduction}

\begin{figure*}[!tbp]
  \centering
\includegraphics[width=\textwidth,clip, trim=0.31cm 2.13cm 2.15cm 2.14cm]{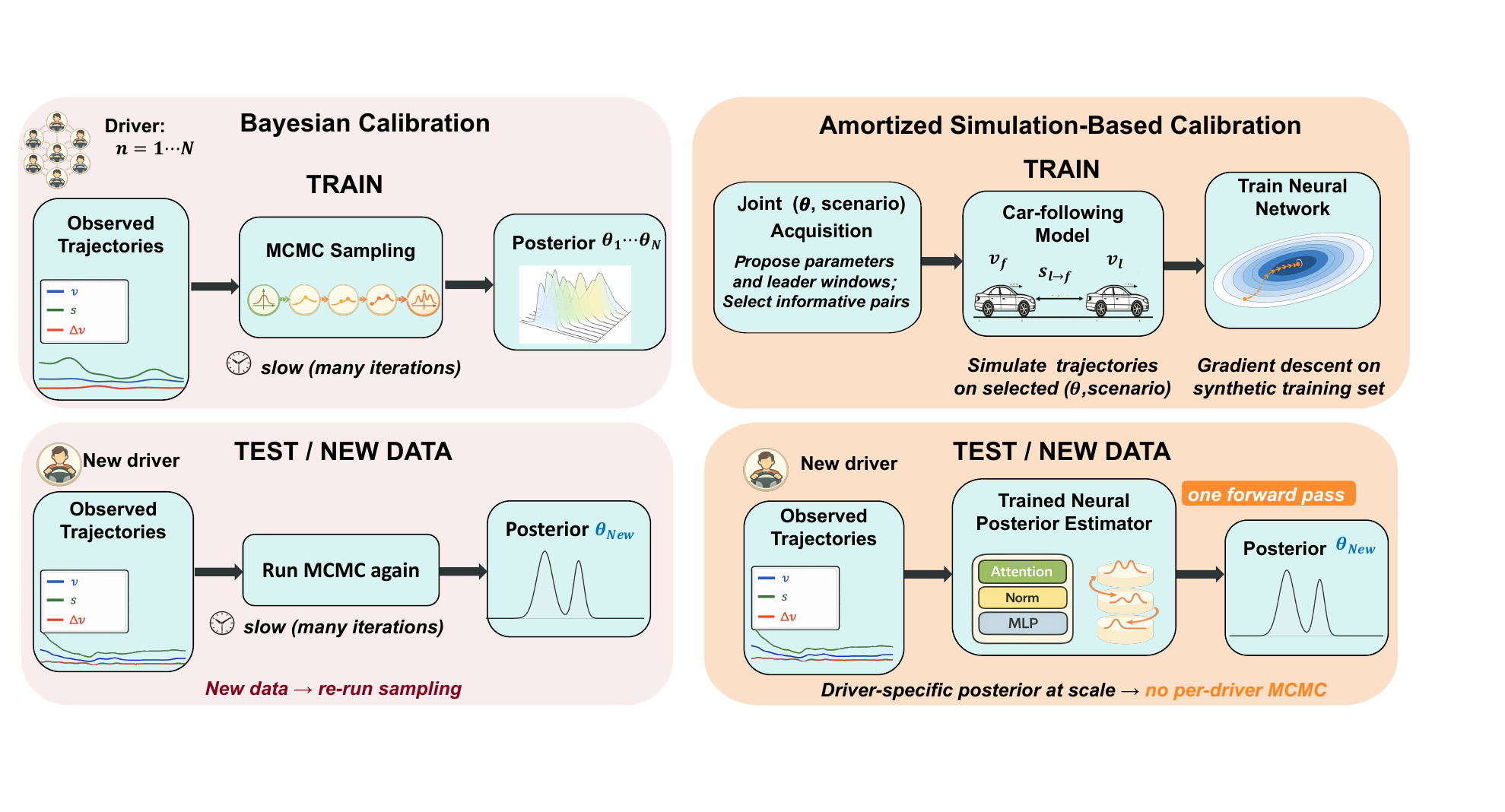}
\caption{Classical per-trajectory Bayesian calibration versus ASBC.
\textbf{Left:} obtain each driver’s posterior by running MCMC per trajectory; re-run when new data arrive.
\textbf{Right:} train once with active joint parameter-scenario acquisition, then infer a driver-specific posterior by one forward pass.}
  \label{fig:bayes_vs_asbc}
\end{figure*}

Human car-following behavior is inherently stochastic and heterogeneous. Even under similar leader trajectories and road conditions, drivers may respond differently, and even the same driver may vary over time due to perception and reaction noise, attention fluctuations, individual driving style, unobserved context, and measurement error in observed trajectories~\cite{zhang2025context,li2020trajectory}. Car-following models provide a compact, parameterized representation of longitudinal interactions and are therefore core building blocks in microscopic traffic simulation~\cite{treiber2000congested}. They underpin safety evaluation for advanced driver-assistance systems (ADAS) and automated vehicles, mixed-traffic analysis, and policy assessment. In these settings, simulation credibility depends not only on matching average behavior but also on capturing uncertainty and tail outcomes across the driver population.

A central calibration challenge is non-identifiability: the same observed trajectory can often be explained by multiple plausible parameter sets~\cite{kesting2008calibrating,punzo2021calibration}. For downstream simulation and decision-making, it is therefore important to quantify parameter uncertainty and propagate it through forward simulation, rather than reporting a single best-fit estimate. However, most practical calibrations still rely on deterministic optimization, in which a goodness-of-fit criterion is chosen and a single parameter vector is obtained by minimizing the discrepancy between simulated and observed trajectories~\cite{keane2021fast,punzo2012can,punzo2021calibration}. In practice, results can be sensitive to modeling choices such as the error metric, time alignment, weighting, and fitting horizon~\cite{punzo2012can,toledo2004statistical}, and the same nonconvex calibration must be repeated for thousands of trajectories when constructing driver populations. These limitations motivate scalable calibration methods that are uncertainty-aware and suitable for risk-sensitive evaluation.

Bayesian calibration addresses these issues by treating the model parameters $\boldsymbol{\theta}$ as random variables and conditioning them on observed driving data $\mathbf{x}$ to obtain a posterior distribution $p(\boldsymbol{\theta}\mid \mathbf{x})$~\cite{zhang2024bayes,zhang2024calibrating}. The posterior enables uncertainty quantification, parameter-correlation analysis, and identifiability diagnostics that are difficult to obtain from point estimates. However, standard Bayesian workflows in these works typically require per-trajectory sampling (e.g., Markov chain Monte Carlo (MCMC)), which becomes computationally prohibitive for modern naturalistic datasets with many drivers and windows. This scalability gap motivates amortized simulation-based inference, where a learned conditional density estimator is trained offline and can then generate posterior samples for new trajectories at test time with negligible additional cost~\cite{papamakarios2016fast,cranmer2020frontier,lueckmann2017flexible}.

To bridge this gap, we develop Active Simulation-Based Calibration (ASBC), which combines amortized posterior inference with active simulation design to make uncertainty-aware car-following calibration practical at scale. As illustrated in Fig.~\ref{fig:bayes_vs_asbc}, ASBC trains a reusable neural posterior estimator offline and avoids per-driver sampling at test time. Given an observed leader--follower trajectory $\mathbf{x}$, ASBC encodes it into a compact context representation and uses a conditional normalizing flow~\cite{winkler2019learning} to efficiently generate posterior samples of $\boldsymbol{\theta}$. To reduce the training simulation budget, ASBC actively acquires informative parameter--scenario pairs by jointly selecting parameter proposals and representative leader-trajectory windows, focusing simulations where the current estimator is most uncertain while remaining realistic.

The proposed framework makes stochastic car-following calibration practical at scale by combining amortized inference with sample-efficient simulation design. It enables driver-specific posteriors that can be carried into downstream simulation, prediction, and risk-oriented evaluation, where system-level conclusions depend on heterogeneity rather than a single best-fit calibration.

Our main contributions are:
\begin{itemize}
  \item \textbf{Scalable, uncertainty-aware calibration:} We propose ASBC, which amortizes Bayesian calibration to infer driver-specific posteriors from observed trajectories with a single forward pass at test time.
  \item \textbf{Joint active simulation design:} We develop a joint acquisition strategy that selects parameter proposals and leader-scenario windows to improve sample efficiency under a fixed simulation budget.
  \item \textbf{Empirical validation:} We evaluate ASBC on the HighD dataset and provide convergence, ablation, robustness, and uncertainty calibration analyses.
\end{itemize}

The remainder of this paper is organized as follows. Section~\ref{sec:related} reviews related work. Section~\ref{sec:method} presents ASBC, including the residual-augmented simulator, amortized posterior inference, and joint active acquisition. Section~\ref{sec:experiments} describes the experimental setup and evaluation protocol. Section~\ref{sec:results} presents the results and diagnostic analyses. Finally, Section~\ref{sec:conclusion} concludes with limitations and future directions.

\section{Related Work}
\label{sec:related}
We situate our contribution at the intersection of (i) car-following model calibration and (ii) amortized simulation-based inference (SBI). Classical calibration has focused on per-trajectory deterministic fitting, while Bayesian variants provide uncertainty quantification but typically remain too expensive to run driver by driver at scale. In parallel, SBI offers fast test-time posterior estimation via learned conditional density models, but standard formulations assume a fixed data-generating context and can still require many simulations. Taken together, prior work still lacks a scalable way to obtain \,\emph{driver-specific posteriors} while \,\emph{actively controlling the simulation budget} and \,\emph{selecting informative driving contexts} (scenarios) rather than treating the scenario as fixed; our framework addresses this by combining amortized Bayesian calibration with active simulation design over a joint parameter--scenario space.

\subsection{Car-Following Model Calibration}
Car-following calibration is commonly posed as an optimization problem: for a given car-following law and observed trajectories, one fits a single parameter vector by minimizing a discrepancy between simulated and observed motion. Results can be sensitive to the chosen loss, noise model, and fitting protocol, even for widely used models such as the Intelligent Driver Model (IDM)~\cite{zhang2024car,kesting2008calibrating}. Many studies therefore emphasize robust search procedures and computational accelerations (e.g., surrogate models or gradient-based speedups) to mitigate nonconvexity and simulator cost~\cite{osorio2019efficient,keane2019fast,ciuffo2013gaussian,keane2021fast}. However, these approaches still deliver point estimates and must be repeated for each driver, limiting scalability and robustness under realistic noise and local optima~\cite{punzo2012can}.

Bayesian calibration replaces point estimates with a posterior over driver parameters. Early work applied MCMC-based stochastic calibration to car-following models~\cite{rahman2013application}, while more recent studies incorporate hierarchical structure and temporally correlated residuals to better capture inter-driver variability and trajectory noise~\cite{zhang2024bayes,zhang2024calibrating}. Likelihood-free Bayesian calibration has also been explored via Approximate Bayesian Computation (ABC) for stochastic and hybrid car-following settings~\cite{jiang2024genericABC,jiang2025abcAV}. Despite improved uncertainty quantification, these Bayesian and likelihood-free methods generally remain computationally intensive because inference is performed trajectory by trajectory and/or requires many simulator evaluations. Prior surrogate-assisted and active-learning ideas reduce cost~\cite{sha2020applying,oladyshkin2020bayesian3}, but they do not fully eliminate per-trajectory inference and typically do not treat scenario selection as a first-class design variable.

\subsection{Amortized Inference and Posterior Estimation}
Amortized inference trains an inference model offline so that posterior estimation for new observations is fast at test time~\cite{zammit2025neural}. SBI instantiates this idea by learning a conditional density model from simulated parameter--data pairs and using it to approximate the posterior without an explicit likelihood~\cite{cranmer2020frontier,boelts2024sbi,deistler2025simulation}. Methods such as SNPE learn flexible posteriors with neural density estimators (including normalizing flows) and can be run sequentially to focus simulations on informative regions of parameter space~\cite{lueckmann2017flexible,greenberg2019automatic,papamakarios2016fast}.

A key practical limitation is simulation cost, motivating active design strategies that adaptively choose simulations to maximize information gain or otherwise improve sample efficiency~\cite{griesemer2024asnpe,beck2016sequential,schultz2018bayesian}. Most active SBI work, however, selects \,\emph{parameters} to simulate under a fixed experimental context. In contrast, car-following data are inherently scenario-dependent: the same driver parameters can express differently under different leader trajectories and traffic regimes. Our approach therefore extends active simulation selection to a \,\emph{joint} parameter--scenario space, enabling efficient training of an amortized posterior estimator that is tailored to car-following calibration.

\section{Methodology}
\label{sec:method}

\begin{figure}[!t]
  \centering
\includegraphics[width=0.5\textwidth]{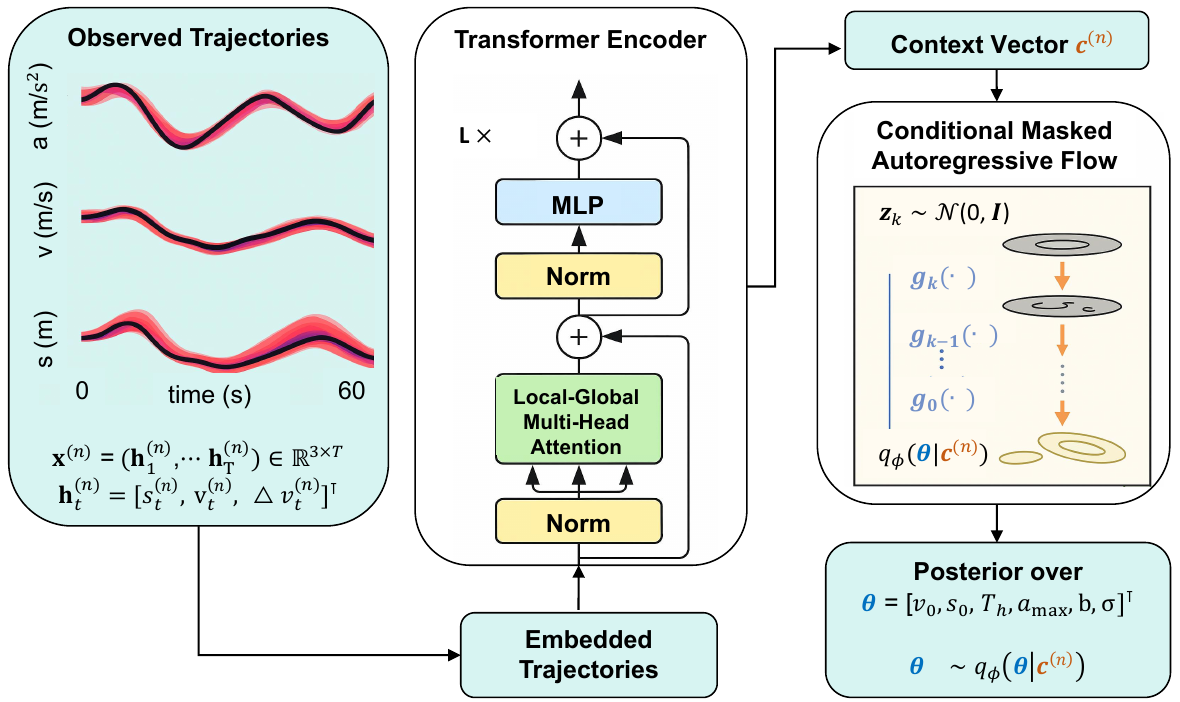}
  \caption{
  Amortized test-time inference module in ASBC.
  The training of this module is simulation-driven and relies on the active learning loop with joint parameter-scenario
  acquisition (Algorithm~\ref{alg:active_amortized_loop}).
  }
  \label{fig:forward_inference}
\end{figure}

This section presents \textbf{ASBC}, an active simulation-based framework for stochastic calibration of car-following models. Given an observed trajectory $\mathbf{x}$, we infer a driver-specific posterior over parameters $\boldsymbol{\theta}$ and draw posterior samples efficiently, so that uncertainty can be propagated through forward simulation to produce stochastic rollouts and distributional safety and performance metrics.

The pipeline forms a closed loop with three components:
(i)~a residual-augmented IDM that defines the calibration model and forward simulator under two residual assumptions (\S\ref{sec:gen_model});
(ii)~a neural posterior estimator that amortizes inference by encoding $\mathbf{x}$ into a context vector and generating posterior samples of $\boldsymbol{\theta}$ with a conditional normalizing flow (\S\ref{sec:npe});
(iii)~an active simulation design module that reduces simulation waste by selecting informative training queries (\S\ref{sec:asnpe}).
Together, these components enable scalable driver-specific posterior sampling across large trajectory datasets. Figure~\ref{fig:forward_inference} focuses on the test-time inference module, while the full training loop with joint parameter–scenario acquisition is summarized in Algorithm~\ref{alg:active_amortized_loop}.

\subsection{Calibration model: residual-augmented IDM}
\label{sec:gen_model}

This section specifies the residual-augmented IDM that serves as our working calibration model and forward simulator. We retain the deterministic IDM core and introduce an additive residual acceleration term to capture unexplained discrepancies between observed and simulated trajectories. This stochastic specification is used for calibration and uncertainty propagation, and it is not meant to imply that real driving trajectories are generated exactly by the model. Given an initial follower state and an exogenous leader input sequence over the observation window, the residual-augmented IDM defines a forward simulator that maps parameters $\boldsymbol{\theta}$ to a trajectory $\mathbf{x}$ through the kinematic updates in \S\ref{sec:kinematics}.

\subsubsection{Notation}
At discrete time step $t$, the follower state is $\mathbf{h}_t = [\,s_t,\; v_t,\; \Delta v_t\,]^\top$,
where $s_t$ is the gap to the leader, $v_t$ is the follower speed, and $\Delta v_t = v_t - v_{\ell,t}$ is the relative speed with respect to the leader speed $v_{\ell,t}$.
We denote the (exogenous) leader input by $\boldsymbol{\ell}_t$, which includes $v_{\ell,t}$ as the driving stimulus, and collect it as $\boldsymbol{\ell}_{\text{lead}}=(\boldsymbol{\ell}_1,\ldots,\boldsymbol{\ell}_T)$ over the observation window. The corresponding observed follower trajectory is $\mathbf{x}=(\mathbf{h}_1,\ldots,\mathbf{h}_T)$, aligned with $\boldsymbol{\ell}_{\text{lead}}$ in time.
Throughout, $\boldsymbol{\ell}_{\text{lead}}$ is treated as known input to the simulator, and $\mathbf{x}$ is obtained by rolling out the state updates from the initial state $\mathbf{h}_1$.

We infer a trajectory-specific posterior over parameters $\boldsymbol{\theta}$.
In both variants, $\boldsymbol{\theta}$ includes the standard IDM parameters and additional residual parameters.

\subsubsection{Deterministic IDM core}
The deterministic IDM acceleration $a_t^{\mathrm{IDM}}$ is defined by \cite{treiber2000congested} as
\begin{align}
a^{\mathrm{IDM}}_t &=
a_{\max}\left[
1-\left(\frac{v_t}{v_0}\right)^{\delta}
-\left(\frac{s^\star(v_t,\Delta v_t)}{s_t}\right)^2
\right], \label{eq:idm}\\
s^\star(v,\Delta v) &=
s_0 + vT + \frac{v\Delta v}{2\sqrt{a_{\max}b}}, \label{eq:sstar}
\end{align}
where $v_0$ is the desired speed, $s_0$ is the jam distance, $T$ is the desired time headway, $a_{\max}$ is the maximum acceleration, $b$ is the comfortable deceleration, and we fix $\delta=4$ following common practice.

\subsubsection{Residual models: i.i.d.\ versus temporally correlated}
Deterministic car-following laws such as the IDM capture average behavior but cannot fully explain naturalistic trajectories. Measurement noise, unobserved driver factors, and model mismatch introduce additional variability, which can also be temporally correlated. Ignoring this unexplained component can make calibration sensitive to the observation window and lead to overly confident uncertainty propagation in downstream simulation. 

Following the modeling spirit of Bayesian IDM (B-IDM) and memory-augmented IDM (MA-IDM)~\cite{zhang2024bayes}, we represent the discrepancy between observed and IDM-predicted accelerations through an additive residual acceleration process,
\begin{equation}
a_t = a_t^{\mathrm{IDM}} + r_t,
\label{eq:acc_total}
\end{equation}
which provides a direct and minimal way to absorb unmodeled effects in the state updates while retaining the IDM core. Building on this residual-augmented formulation, we develop an amortized inference and active simulation design pipeline for scalable driver-specific posterior sampling. We consider two residual assumptions: an i.i.d.\ Gaussian baseline and a Mat\'ern-$5/2$ temporally correlated alternative used for sensitivity analysis.

\paragraph{ASBC-Gaussian (i.i.d.\ residual acceleration)}
We assume i.i.d.\ residual acceleration
\begin{equation}
r_t \equiv \varepsilon_t,\qquad \varepsilon_t \sim \mathcal{N}(0,\sigma^2),\quad \text{i.i.d.\ across } t.
\label{eq:gauss_res}
\end{equation}
This simple baseline is efficient to simulate and matches the i.i.d.\ residual assumption commonly used in B-IDM~\cite{zhang2024bayes}.

\paragraph{ASBC-Mat\'ern (Mat\'ern-$5/2$ correlated residual acceleration)}
To reflect temporally dependent discrepancies, we also consider a correlated residual model. We model $r(t)$ as a zero-mean GP with a Mat\'ern-$5/2$ kernel,
\begin{equation}
r(t) \sim \mathcal{GP}\!\left(0,\; k_{\mathrm{M52}}(t,t';\sigma,\ell)\right),
\label{eq:gp_prior}
\end{equation}
where
{\footnotesize
\begin{equation}
k_{\mathrm{M52}}(t,t') = \sigma^2\left(1+\frac{\sqrt{5}\,|t-t'|}{\ell} + \frac{5(t-t')^2}{3\ell^2}\right)
\exp\!\left(-\frac{\sqrt{5}\,|t-t'|}{\ell}\right),
\label{eq:matern52}
\end{equation}
}
and $\ell$ is measured in seconds. This option follows MA-IDM~\cite{zhang2024bayes} in using correlated residuals to capture serial dependence in the discrepancy.

\subsubsection{Stochastic kinematics}
\label{sec:kinematics}
Given leader speed $v_{\ell,t}$ and acceleration $a_t$ from \eqref{eq:acc_total}, we update the follower speed and gap by
\begin{align}
v_{t+1} &= v_t + a_t\,\Delta t, \label{eq:vel_update}\\
s_{t+1} &= s_t + (v_{\ell,t}-v_t)\,\Delta t - \tfrac{1}{2}a_t\,\Delta t^2. \label{eq:gap_update}
\end{align}
These updates define a fast forward simulator under either residual assumption: given $\boldsymbol{\theta}$, the leader input sequence, and the initial state, iterating \eqref{eq:acc_total}--\eqref{eq:gap_update} produces the full trajectory $\mathbf{x}$.

\subsubsection{Priors for IDM parameters and residual hyperparameters}
\label{sec:prior}

We place weakly informative priors in log space to enforce positivity, following the  practice in~\cite{zhang2024bayes}. Let the deterministic IDM parameters be
$\boldsymbol{\theta}_{\mathrm{IDM}}=[v_0,s_0,T,a_{\max},b]^\top$.
We use a diagonal log-normal prior in log space centered at the commonly recommended IDM values $\boldsymbol{\theta}_{\mathrm{rec}}=[33.3,\,2.0,\,1.6,\,1.5,\,1.67]$~\cite{treiber2012trajectory},
with a broad variance to cover realistic inter-driver heterogeneity; we also restrict the support to physically plausible ranges to avoid unstable simulations. In practice, this truncation is implemented during simulation by rejecting draws outside the feasible ranges.

To model unexplained acceleration variability, we introduce residual hyperparameters and include them in the full parameter vector $\boldsymbol{\theta}$.
For ASBC-Gaussian, we use
\[
\boldsymbol{\theta} = [\,\boldsymbol{\theta}_{\mathrm{IDM}},\; \sigma\,]^\top,
\]
where $\sigma$ is the residual scale and follows a log-normal prior.
For ASBC-Mat\'ern, we use
\[
\boldsymbol{\theta} = [\,\boldsymbol{\theta}_{\mathrm{IDM}},\; \sigma,\; \ell\,]^\top,
\]
where $\ell$ is the correlation length scale (in seconds); we place log-normal priors on both $\sigma$ and $\ell$.
The specific numerical settings for the prior configuration and truncation ranges are provided in the Appendix~\ref{app:priors}.

\begin{table}[t]
\centering
\caption{Summary of notations used for data, encoding, and inference.}
\begin{tabularx}{\columnwidth}{@{}l >{\raggedright\arraybackslash}X@{}}
\toprule
\textbf{Symbol} & \textbf{Description} \\
\midrule
$\mathbf{h}_t \in \mathbb{R}^3$ 
& Follower state at time $t$: gap $s_t$, speed $v_t$, and relative speed $\Delta v_t$ \\
$\boldsymbol{\ell}_t$ 
& Exogenous leader input at time $t$, used to define the driving context \\
$\mathbf{x}=(\mathbf{h}_1,\ldots,\mathbf{h}_T)$ 
& Observed follower trajectory of length $T$ \\
$\boldsymbol{\ell}_{\text{lead}}=(\boldsymbol{\ell}_1,\ldots,\boldsymbol{\ell}_T)$ 
& Leader trajectory aligned with $\mathbf{x}$, treated as known input to the simulator \\
$\mathcal{X}=\{\mathbf{x}^{(n)}\}_{n=1}^N$ 
& Dataset of $N$ observed follower trajectories \\
$\mathcal{L}=\{\boldsymbol{\ell}^{(n)}\}_{n=1}^N$ 
& Dataset of full-length exogenous leader trajectories aligned with $\mathcal{X}$ \\
$L$ 
& Leader trajectory window extracted from $\boldsymbol{\ell}^{(n)}$, used as a scenario context in active acquisition \\
$\boldsymbol{\theta}^{(n)}$ 
& Driver-specific latent parameters for trajectory $n$; dimension depends on the residual model \\
$f_{\psi}(\cdot)$ 
& Trajectory encoder mapping an observed trajectory to a context vector $\mathbf{c}\in\mathbb{R}^d$ \\
$\mathbf{c}^{(n)}=f_{\psi}(\mathbf{x}^{(n)})$ 
& Context embedding for trajectory $n$ \\
$q_{\phi}(\boldsymbol{\theta} \mid \mathbf{x})$ 
& Amortized posterior estimator of $\boldsymbol{\theta}$ conditioned on observations \\
\bottomrule
\end{tabularx}
\label{tab:notation}
\end{table}

\subsection{Neural posterior estimation (NPE)}
\label{sec:npe}

Given the residual-augmented calibration model in \S\ref{sec:gen_model}, our goal is to infer a trajectory-specific posterior over parameters from an observed car-following trajectory. Specifically, for each observation $\mathbf{x}^{(n)}$, we aim to approximate the posterior $p(\boldsymbol{\theta}^{(n)}\mid \mathbf{x}^{(n)})$ with a single amortized model $q_{\phi}(\boldsymbol{\theta}\mid \mathbf{x})$. Once trained, this estimator can generate posterior samples for a new trajectory by a fast forward pass, avoiding per-trajectory sampling or optimization.

Related Bayesian calibration studies consider pooled and unpooled formulations~\cite{zhang2024bayes}. The pooled formulation learns a single posterior using all car-following pairs, which is statistically efficient but can blur individual heterogeneity. The unpooled formulation performs per-driver or per-pair inference, which better aligns with individualized behavior but can be costly when repeated across large datasets. Our amortized posterior estimator targets the unpooled use case while sharing statistical strength across the dataset through a single trained inference network.

Our estimator has two stages. A trajectory encoder $f_{\psi}$ maps each observed trajectory to a fixed-dimensional context representation. Conditioned on this context, a conditional normalizing flow models the posterior $q_{\phi}(\boldsymbol{\theta}\mid \mathbf{c})$ and supports efficient sampling for inference and acquisition in \S\ref{sec:asnpe}. Table~\ref{tab:notation} lists the main symbols used in the method.

\subsubsection{Trajectory encoder $f_{\psi}$}
\label{sec:encoder}

To support amortized inference with variable-length observations, we encode each follower trajectory $\mathbf{x}\in\mathbb{R}^{T\times 3}$ into a fixed-dimensional context vector
\[
\mathbf{c}=f_{\psi}(\mathbf{x})\in\mathbb{R}^{d}.
\]
We implement $f_{\psi}$ using a Transformer-style sequence encoder~\cite{vaswani2017attention,xiong2021nystromformer} that captures both local temporal patterns and longer-range dependencies over the observation window. The resulting context vector $\mathbf{c}$ summarizes trajectory-level response features and serves as the conditioning input to the posterior flow in \S\ref{sec:maf}. Details of the encoder architecture are provided in the Appendix~\ref{app:encoder_details}.

\subsubsection{Conditional masked autoregressive flow}
\label{sec:maf}

We model the trajectory-conditioned posterior with a conditional masked autoregressive flow (MAF)~\cite{papamakarios2017masked}. Given a context vector $\mathbf{c}$, the flow defines a flexible conditional density over an unconstrained latent vector $\mathbf{u}\in\mathbb{R}^{D}$, denoted by $q_{\phi}(\mathbf{u}\mid \mathbf{c})$. Sampling is efficient: we draw a base Gaussian noise vector and apply the inverse flow to obtain $\mathbf{u}$ conditioned on $\mathbf{c}$. The same model also supports exact evaluation of $\log q_{\phi}(\mathbf{u}\mid \mathbf{c})$ via the standard change-of-variables rule.

To enforce physical constraints such as positivity of IDM parameters and residual scales, we map $\mathbf{u}$ to the physical parameter vector $\boldsymbol{\theta}$ using an element-wise bijection
\[
\boldsymbol{\theta}=T(\mathbf{u})=\mathrm{softplus}(\mathbf{u})+\varepsilon,\qquad \varepsilon>0.
\]
Accordingly, posterior samples are obtained by $\mathbf{u}\sim q_{\phi}(\cdot\mid \mathbf{c})$ followed by $\boldsymbol{\theta}=T(\mathbf{u})$. When needed, the physical-space density is computed by
{\small
\[
\log q_{\phi}(\boldsymbol{\theta}\mid \mathbf{c})
=
\log q_{\phi}(\mathbf{u}\mid \mathbf{c})
-
\log\left|\det\!\left(\tfrac{\partial T(\mathbf{u})}{\partial \mathbf{u}}\right)\right|,
\qquad \mathbf{u}=T^{-1}(\boldsymbol{\theta})
\]
}
with the Jacobian term being diagonal for the softplus transform. Full expressions for the flow likelihood and the transform details are provided in the Appendix~\ref{app:flow_transform}.

\paragraph*{Epistemic uncertainty via Monte Carlo (MC) dropout}
We use MC dropout to approximate epistemic uncertainty in the learned posterior model. By enabling dropout at inference time, we obtain an ensemble of posterior evaluations that is used in the acquisition function in \S\ref{sec:asnpe} to identify informative parameter--scenario queries.

\subsubsection{Simulation-driven training objective}
\label{sec:objective}

\paragraph*{Simulation-based training principle}
In general, amortized posterior estimation is trained from simulated calibration pairs generated by a forward simulator. Given a parameter vector $\boldsymbol{\theta}$ and an exogenous scenario context $L$, defined as a short window extracted from an observed leader trajectory $\boldsymbol{\ell}_{\text{lead}}$, the simulator produces a synthetic follower trajectory,
\[
\boldsymbol{\theta} \sim \tilde p(\boldsymbol{\theta}), \qquad \mathbf{x}_{\text{sim}} = \mathrm{Sim}(\boldsymbol{\theta}, L),
\]
where $\mathrm{Sim}(\cdot)$ denotes the residual-augmented IDM forward model in \S\ref{sec:gen_model} and $\tilde p(\boldsymbol{\theta})$ is a proposal distribution used for simulation. Each simulated trajectory is mapped to a context embedding $\mathbf{c}=f_{\psi}(\mathbf{x}_{\text{sim}})$, and a conditional density model $q_{\phi}(\boldsymbol{\theta}\mid \mathbf{c})$ is fit by maximizing the conditional log-likelihood
\begin{equation}
\max_{\psi,\phi}\;
\mathbb{E}\Big[
\log q_{\phi}\!\big(\boldsymbol{\theta} \mid f_{\psi}(\mathbf{x}_{\text{sim}})\big)
\Big],
\label{eq:npe_principle}
\end{equation}
where the expectation is taken over the simulation procedure that generates $(\boldsymbol{\theta},\mathbf{x}_{\text{sim}})$ pairs.

\paragraph*{Sequential training}
Rather than training from a fixed proposal distribution, we adopt a SNPE procedure~\cite{papamakarios2016fast}. Across rounds, new simulated training pairs are generated by the active acquisition module in \S\ref{sec:asnpe} and added to a fixed-capacity buffer, after which the encoder and flow are updated by maximizing \eqref{eq:npe_loss_final} on minibatches from the buffer.

\paragraph*{Maximum-likelihood training}
We train the encoder and flow by maximizing the conditional log-likelihood on buffered simulated pairs:
\begin{equation}
\mathcal{L}(\psi,\phi)
=
\mathbb{E}_{(\boldsymbol{\theta},\mathbf{c}) \sim \mathcal{D}_r}
\left[
\log q_{\phi}(\boldsymbol{\theta} \mid \mathbf{c})
\right],
\label{eq:npe_loss_final}
\end{equation}
where $\mathcal{D}_r$ denotes the empirical distribution induced by the round-$r$ acquisition and simulation procedure (see \S \ref{sec:asnpe}). This objective is standard in NPE and can be related to fitting the posterior approximation in the chosen model family~\cite{papamakarios2016fast}. 

\paragraph*{Optimization}
The objective in \eqref{eq:npe_loss_final} is differentiable with respect to both encoder parameters $\psi$ and flow parameters $\phi$. We optimize it using stochastic gradient methods over minibatches sampled from the buffer.

\subsection{Active learning with joint acquisition}
\label{sec:asnpe}

This section describes the active-loop used to train the amortized posterior estimator efficiently. The practical goal is to spend simulation budget where it improves driver-specific calibration the most. Each round proposes candidate parameters, pairs them with diverse leader-window scenarios, selects informative parameter--scenario pairs, simulates follower trajectories under the residual-augmented IDM in \S\ref{sec:gen_model}, and updates the encoder and posterior flow using the simulation-driven objective \eqref{eq:npe_loss_final} in \S\ref{sec:objective}.

\paragraph{Overview of the loop}
The loop follows a four-step structure: proposal update, joint acquisition, simulation and encoding, and posterior refinement. In contrast to selecting parameters alone, we actively select pairs $(\boldsymbol{\theta},L)$ so that parameters are queried under informative and representative driving contexts. A fixed-capacity first-in first-out (FIFO) buffer stores recent training pairs to stabilize retraining across rounds.

\paragraph{Soft-shrink proposal over parameters}
At round $r=0$, parameters are drawn from the prior $p(\boldsymbol{\theta})$ to ensure broad coverage. For $r\ge 1$, we define a soft-shrink mixture proposal
\begin{equation}
\label{eq:soft_proposal_new}
\tilde p_r(\boldsymbol{\theta}) \;=\; \lambda_r\,p(\boldsymbol{\theta}) \;+\; (1-\lambda_r)\,\tilde q_{r-1}(\boldsymbol{\theta}),
\qquad \lambda_r\in[0,1],
\end{equation}
where $\tilde q_{r-1}(\boldsymbol{\theta})$ denotes a posterior-informed proposal constructed from the current amortized model evaluated on observed trajectories.\footnote{In practice, $\tilde q_{r-1}$ is constructed by drawing posterior samples from $q_{\phi_{r-1}}(\boldsymbol{\theta}\mid \mathbf{x}^{(n)})$ for a subset of observed trajectories and pooling these samples to form an empirical proposal distribution. Conceptually, this corresponds to mixing the per-trajectory posteriors over the selected subset of trajectories.}

Annealing $\lambda_r$ toward smaller values gradually shifts from exploration to exploitation while preventing early over-concentration.

\paragraph{Leader bank and representativeness}
We maintain a leader-window bank to cover diverse driving contexts. Real leader windows in $\mathcal{L}_{\mathrm{real}}$ are extracted from the observed leader trajectories $\mathcal{L}$ using a fixed window length and stride. At round $r$, leader windows are sampled from a mixture
\begin{equation}
\label{eq:leader_mix_new}
\mathcal{L}_r \;=\; (1-\alpha_r)\,\mathcal{L}_{\mathrm{real}} \;+\; \alpha_r\,\mathcal{L}_{\mathrm{syn}},
\qquad \alpha_r\in[0,1],
\end{equation}
where $\mathcal{L}_{\mathrm{real}}$ consists of windows extracted from observed leader trajectories and $\mathcal{L}_{\mathrm{syn}}$ consists of lightly augmented windows that preserve basic kinematic plausibility.\footnote{We use low-amplitude perturbations and simple time rescaling/warping with physics-based filtering; details are provided in the Appendix~\ref{app:leader_aug}.}
Each leader window $L$ is mapped to a fixed feature vector $\varphi(L)\in\mathbb{R}^{d_\ell}$ computed from the leader speed and acceleration profiles over the window (e.g., concatenating simple summary statistics such as mean, standard deviation, extrema, and average absolute first differences). To encourage coverage, we define a representativeness score
{\small
\begin{equation}
\label{eq:leader_repr_new}
\rho(L)\;=\;\frac{1}{\varepsilon+\frac{1}{K}\sum_{k=1}^{K}\bigl\|\varphi(L)-\varphi(L_{(k)})\bigr\|_2}\,,
\qquad \varepsilon>0,
\end{equation}
}
where $L_{(k)}$ are the $K$ nearest leader windows to $L$ in feature space.

\paragraph{Joint acquisition over parameter--scenario pairs}
At round $r$, we draw parameter candidates $\{\boldsymbol{\theta}_j\}_{j=1}^{N}\sim \tilde p_r$ and leader candidates $\{L_k\}_{k=1}^{N_\ell}\sim \mathcal{L}_r$, and form a candidate pool $\mathcal{C}_r=\{(\boldsymbol{\theta}_j,L_k)\}$. We quantify parameter uncertainty using an ASNPE-style \cite{griesemer2024asnpe} score $\alpha_r(\boldsymbol{\theta})$ computed from MC-dropout evaluations of the posterior model.
Intuitively, $\alpha_r(\boldsymbol{\theta})$ is large when the current posterior model assigns unstable or highly variable density to $\boldsymbol{\theta}$ under dropout, indicating epistemic uncertainty. We then select pairs using a greedy strategy. Let $\mathcal{S}$ denote the set of parameter--scenario pairs that have already been selected in the current round, initialized as $\mathcal{S}=\varnothing$ and updated as pairs are added one by one. We define a joint acquisition score
{\small
\begin{equation}
\label{eq:joint_score_new}
s_r(\boldsymbol{\theta},L)
=\alpha_r(\boldsymbol{\theta})\,\rho(L)
-\gamma\,\mathrm{DivPen}\!\bigl(\mathcal{S}\cup\{(\boldsymbol{\theta},L)\}\bigr),
\qquad \gamma\ge 0,
\end{equation}
}
where $\mathrm{DivPen}(\cdot)$ is a diversity penalty that discourages selecting near-duplicate pairs in the joint space.\footnote{We implement $\mathrm{DivPen}$ using a kernel dispersion term over leader features $\varphi(L)$ and a rescaled distance in parameter space.}
Greedy maximization of $s_r(\boldsymbol{\theta},L)$ yields a top-$B$ set of pairs that are simultaneously high-uncertainty, representative, and diverse.

\paragraph{Simulation, buffering, and posterior refinement}
For each selected pair $(\boldsymbol{\theta}_b,L_b)$, we simulate a follower trajectory $\mathbf{x}_b=\mathrm{Sim}(\boldsymbol{\theta}_b,L_b)$ under the residual-augmented IDM in \S\ref{sec:gen_model}, encode it into a context vector $\mathbf{c}_b=f_{\psi}(\mathbf{x}_b)$, and append $(\boldsymbol{\theta}_b,\mathbf{c}_b)$ to a fixed-capacity FIFO buffer $\mathcal{D}_r$. After appending the new batch, the oldest samples are discarded to keep the buffer size fixed, so $\mathcal{D}_r$ induces an empirical training distribution over recent simulated pairs. We then update the encoder and posterior flow by maximizing ~\eqref{eq:npe_loss_final} on minibatches sampled from $\mathcal{D}_r$.

\paragraph{Monitoring and stopping}
To assess convergence and prevent unnecessary simulation, we monitor validation metrics computed on held-out observed trajectories and apply early stopping across rounds when improvements become marginal. This provides a practical diagnostic for stable training and reproducible comparisons across datasets. The overall procedure is summarized in Algorithm~\ref{alg:active_amortized_loop}.

\begin{algorithm}[!t]
\caption{ASBC training with joint parameter--scenario acquisition}
\label{alg:active_amortized_loop}
\begin{algorithmic}[1]
\State \textbf{Input:} prior $p(\boldsymbol{\theta})$; observed data $(\mathcal{X},\mathcal{L})$; simulator $\mathrm{Sim}(\cdot)$; rounds $R$; pairs per round $B$; buffer capacity $N_{\mathrm{buf}}$\;
\State \textbf{Initialize:} FIFO buffer $\mathcal{D}_0\leftarrow\varnothing$; initialize encoder and posterior estimator $(f_{\psi},q_{\phi})$\;

\Statex \textbf{Warm-up (prior simulations)}\;
\For{$b=1$ to $B_0$}
  \State sample $\boldsymbol{\theta}_b\sim p(\boldsymbol{\theta})$ and a real leader window $L_b$\;
  \State simulate $\mathbf{x}_b\leftarrow \mathrm{Sim}(\boldsymbol{\theta}_b,L_b)$ and encode $\mathbf{c}_b\leftarrow f_{\psi}(\mathbf{x}_b)$\;
  \State push $(\boldsymbol{\theta}_b,\mathbf{c}_b)$ into $\mathcal{D}_0$; if $|\mathcal{D}_0|>N_{\mathrm{buf}}$, discard the oldest samples\;
\EndFor
\State train $(f_{\psi},q_{\phi})$ on minibatches from $\mathcal{D}_0$ via Eq.~\eqref{eq:npe_loss_final}\;

\Statex \textbf{Active refinement}\;
\For{$r=1$ to $R$}
  \State update proposal $\tilde p_r(\boldsymbol{\theta})$ (Eq.~\eqref{eq:soft_proposal_new})\;
  \State build leader-window candidates $\mathcal{L}_r$ and scores $\rho(L)$ (Eqs.~\eqref{eq:leader_mix_new}--\eqref{eq:leader_repr_new})\;
  \State select $B$ pairs $\{(\boldsymbol{\theta}_b,L_b)\}_{b=1}^B$ by maximizing Eq.~\eqref{eq:joint_score_new}\;
  \For{$b=1$ to $B$}
    \State simulate $\mathbf{x}_b\leftarrow \mathrm{Sim}(\boldsymbol{\theta}_b,L_b)$ and encode $\mathbf{c}_b\leftarrow f_{\psi}(\mathbf{x}_b)$\;
    \State push $(\boldsymbol{\theta}_b,\mathbf{c}_b)$ into $\mathcal{D}_r$; if $|\mathcal{D}_r|>N_{\mathrm{buf}}$, discard the oldest samples\;
  \EndFor
  \State update $(f_{\psi},q_{\phi})$ on minibatches from $\mathcal{D}_r$ via Eq.~\eqref{eq:npe_loss_final}\;
  \State optionally monitor validation metrics and early stop\;
\EndFor

\State \textbf{Return:} trained $(f_{\psi},q_{\phi})$ enabling fast sampling $\boldsymbol{\theta}\sim q_{\phi}(\boldsymbol{\theta}\mid f_{\psi}(\mathbf{x}))$\;
\end{algorithmic}
\end{algorithm}

\section{Experiments}
\label{sec:experiments}

This section evaluates ASBC and addresses three questions:
(i)~How does ASBC compare with established calibration baselines in simulation fidelity and uncertainty quantification?
(ii)~How sensitive are results to the residual model (i.i.d.\ Gaussian versus temporally correlated Mat\'ern-$5/2$)?
(iii)~Does the active learning loop improve validation performance in a stable and sample-efficient manner, and does it stop when gains become marginal?

\subsection{Dataset and preprocessing}
\label{sec:data}

\subsubsection{Dataset}
\label{sec:dataset}
We evaluate on the HighD dataset~\cite{krajewski2018highd}, which provides naturalistic highway trajectories with predominantly lane-keeping car-following interactions. We downsample trajectories from 25~Hz to 5~Hz (retaining every fifth frame), yielding a simulation time step $\Delta t=0.2$~s and reducing the cost of repeated simulation and training. This setting follows common practice in Bayesian calibration of IDM-type models~\cite{zhang2024bayes}.

\subsubsection{Leader--follower segments}
\label{sec:preprocess}
We extract leader--follower segments by retaining only intervals in which the follower continuously tracks the same leader; a leader change initiates a new segment. Within each segment, we construct the follower state $\mathbf{h}_t=[s_t,\,v_t,\,\Delta v_t]^\top$ together with the time-aligned leader input required by the simulator. To obtain a fixed-size context for the encoder, we extract one window of duration $T_E$ from each follower vehicle and use it as the encoder input for posterior inference, yielding a driver-specific posterior. Unless otherwise noted, we set $T_E=15$~s based on the encoder history-length sweep (Figure~\ref{fig:highd_history_sweep}), which achieves near-saturated performance while keeping training time manageable. Finally, we discard segments shorter than $T_{\min}=40$~s to ensure sufficient temporal context for calibration.

\subsubsection{Train/validation/test split}
\label{sec:split}
We split segments into training/validation/test sets with a 60\%/10\%/30\% ratio, stratified by follower vehicle IDs. This ID-based split reduces leakage across splits and evaluates generalization to unseen drivers.

\subsection{Evaluation protocol and metrics}
\label{sec:eval_protocol}

To keep all tables and figures directly comparable, we evaluate each model variant under a standardized protocol. In all evaluations, the leader trajectory is treated as a known exogenous input, and uncertainty is propagated by drawing parameter samples from the learned posterior and rolling out the residual-augmented IDM simulator under the corresponding residual assumption. Unless otherwise stated, all quantitative results are computed on the held-out test split; validation is used only for monitoring and early stopping.

\paragraph{Windowed short-horizon simulation}
For each held-out leader--follower pair, we evaluate predictive performance on sliding windows of fixed horizon $H$ with stride $S$, using up to $m$ windows per pair to control computation. In our HighD experiments, $\Delta t=0.2$~s (5~Hz), and we use $H=50$ and $S=20$, corresponding to 10~s windows with a 4~s stride. For each window, we reset the simulator to the observed follower state at the window start and roll out $H$ steps using the aligned leader segment. This reset-at-window-start design focuses the evaluation on short-horizon predictive fidelity and uncertainty, without letting long-horizon drift dominate the score.

\paragraph{Point accuracy metric}
For each state variable $x\in\{s,v,a\}$, we compute the root mean squared error (RMSE) between the ground truth and the posterior predictive mean:
\begin{equation}
\widehat{x}_t=\frac{1}{n}\sum_{i=1}^n x^{(i)}_t,
\qquad
\mathrm{RMSE}(x)=\sqrt{\frac{1}{H}\sum_{t=1}^H\left(\widehat{x}_t-x_t\right)^2}.
\end{equation}

\paragraph{Distributional quality metric}
To evaluate the full predictive distribution, we use the Energy Score (ES), a proper scoring rule for sample-based forecasts. For a trajectory segment represented as a vector $y\in\mathbb{R}^{d}$ and predictive samples $\{y^{(i)}\}_{i=1}^n$,
\begin{equation}
\mathrm{ES}(y)=\frac{1}{n}\sum_{i=1}^n \|y^{(i)}-y\|_2
-\frac{1}{2n^2}\sum_{i=1}^n\sum_{j=1}^n \|y^{(i)}-y^{(j)}\|_2.
\end{equation}
In practice, we compute ES separately for $s$, $v$, and $a$ by treating each $H$-step time series as a vector (and we also report an averaged ES where appropriate). Lower ES indicates a sharper and better-aligned predictive distribution.\subsection{Reproducibility and implementation details}
\label{sec:repro}

We report the key settings needed to reproduce our experiments; a complete hyperparameter table and additional diagnostics are provided in the Appendix~\ref{app:impl_hparams}. The full codebase will be made publicly available upon publication.

\paragraph{Training and active loop}
We train ASBC with up to $R=10$ active rounds with cross-round early stopping. Round 0 performs prior-simulation warm-up with $B_0=4000$ draws. Each active round selects and simulates $B=2000$ $(\boldsymbol{\theta},L)$ pairs and then fine-tunes the encoder and conditional flow for at most 100 epochs using Adam (\texttt{lr}$=10^{-3}$, \texttt{batch\_size}$=128$) on a fixed-capacity FIFO replay buffer. Detailed schedules (including proposal and leader-bank mixing) are provided in the Appendix~\ref{app:proposal}.

\paragraph{Early stopping and evaluation configuration}
To avoid spending simulation budget after saturation, we apply round-level early stopping based on validation performance. Since real trajectories do not provide ground-truth parameters, validation is performed on a fixed synthetic hold-out set of simulated $(\boldsymbol{\theta},\mathbf{x}_{\mathrm{sim}})$ pairs generated once with a fixed random seed using the same simulator and leader-window construction as training. Unless otherwise stated, evaluation uses $n=500$ posterior samples per pair.

\paragraph{Data usage}
The posterior estimator is trained on simulated pairs $(\boldsymbol{\theta},\mathbf{x}_{\mathrm{sim}})$ produced by the residual-augmented simulator under leader-window inputs. Real trajectories are used only to construct the posterior-based proposal distribution in the active loop and to compute validation/testing metrics.

\begin{figure*}[!t]
\centering
\includegraphics{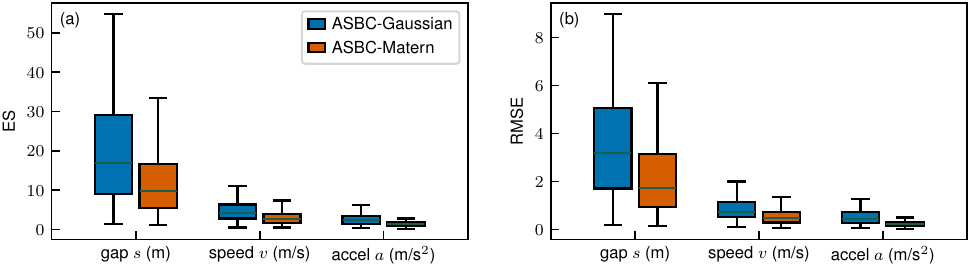}
\caption{Per-window 10-second simulation error distributions on HighD for ASBC. (a) ES and (b) RMSE over gap $s$ (m), speed $v$ (m/s), and acceleration $a$ (m/s$^2$). Colors indicate ASBC-Gaussian and ASBC-Mat\'ern. Boxes show the interquartile range and the median.}
\label{fig:highd_errdist}
\end{figure*}

\begin{table*}[!ht]
\centering
\caption{Evaluation of 10-second stochastic simulations.}
\label{tab:highd_summary}
\small
\setlength{\tabcolsep}{5pt}
\begin{tabular*}{\linewidth}{@{\extracolsep{\fill}}lccc|ccc}
\toprule
\multirow{2}{*}{Method} & \multicolumn{3}{c}{RMSE $\downarrow$} & \multicolumn{3}{c}{ES $\downarrow$} \\
\cmidrule(lr){2-4}\cmidrule(lr){5-7}
 & $s$ & $v$ & $a$ & $s$ & $v$ & $a$ \\
\midrule
ASBC-Gaussian
& $2.633 \pm 0.150$ & $0.627 \pm 0.035$ & $0.291 \pm 0.018$
& $15.510 \pm 1.141$ & $3.571 \pm 0.248$ & $1.590 \pm 0.093$ \\
ASBC-Mat\'ern
& $2.455 \pm 0.209$ & $0.628 \pm 0.077$ & $0.302 \pm 0.031$
& $13.452 \pm 0.489$ & $3.367 \pm 0.278$ & $1.511 \pm 0.179$ \\
\midrule
Pooled MA-IDM
& $3.381$ & $1.088$ & $0.399$
& $24.591$ & $6.340$ & $2.378$ \\
Pooled B-IDM
& $3.957$ & $1.044$ & $0.439$
& $26.709$ & $7.033$ & $2.744$ \\
\bottomrule
\end{tabular*}
\end{table*}

\subsection{Methods compared}
\label{sec:methods_compared}

We compare two variants of ASBC against classical Bayesian IDM calibration baselines. This comparison isolates (i) the benefit of amortized inference, where a single learned posterior estimator is reused across trajectories, and (ii) the impact of the residual noise structure (i.i.d.\ versus temporally correlated errors).

\subsubsection{Our methods}
\label{sec:methods_ours}

\textbf{ASBC-Gaussian.}
Residual-augmented IDM with additive i.i.d.\ Gaussian acceleration noise, trained with the proposed amortized inference and active design pipeline.

\textbf{ASBC-Mat\'ern.}
Same pipeline, but the residual acceleration follows a temporally correlated Mat\'ern-$5/2$ process to capture persistent model mismatch.

\subsubsection{Baselines}
\label{sec:methods_baselines}

\textbf{B-IDM (pooled).}
Bayesian IDM with i.i.d.\ Gaussian acceleration noise. We use a pooled formulation (one global parameter vector shared across drivers) to keep dataset-level evaluation tractable.

\textbf{MA-IDM (pooled).}
Memory-augmented IDM of~\cite{zhang2024bayes}, which models temporally correlated residuals and improves realism on HighD relative to B-IDM. We also use a pooled formulation, since unpooled Bayesian calibration would require running MCMC separately for each trajectory/driver and is computationally expensive at scale.

For both baselines, we use the open-sourced implementation (\url{https://github.com/Chengyuan-Zhang/IDM_Bayesian_Calibration.git}). Predictive simulations are generated by sampling $\boldsymbol{\theta}$ from the pooled posterior and rolling out the corresponding residual-augmented IDM under held-out leader inputs, using the same evaluation protocol as ASBC. We report a single result for these baselines because the posterior is fixed once calibrated; repeating runs would mainly reflect Monte Carlo sampling noise rather than variability across random seeds and initializations.

\section{Results and Discussion}
\label{sec:results}
\subsection{Quantitative results}
We begin with a windowed short-horizon evaluation on the HighD test split, following the protocol in Section~\ref{sec:eval_protocol}. We report both point accuracy (RMSE) and distributional quality (Energy Score, ES) on 10\,s rollouts under held-out leader inputs. Table~\ref{tab:highd_summary} summarizes performance for our ASBC variants and the pooled Bayesian calibration baselines. For ASBC, values are reported as mean$\pm$standard deviation over five random seeds. Overall, both ASBC variants substantially outperform the pooled baselines across RMSE and ES, indicating that context-conditioned (driver--scenario) calibration is important for achieving accurate and uncertainty-aware short-horizon simulations at scale.

We next isolate the effect of the residual model by comparing ASBC-Gaussian and ASBC-Mat\'ern. ASBC-Mat\'ern consistently improves distributional fidelity, reducing ES across all three variables ($s$, $v$, $a$), with the largest gain on spacing $s$ (15.51$\rightarrow$13.45). In terms of point accuracy, the Mat\'ern model improves spacing RMSE (2.63$\rightarrow$2.46) while yielding comparable RMSE for speed and acceleration. This pattern suggests that modeling temporally correlated disturbances primarily improves the match of the predictive distribution, even when the predictive mean changes only modestly.

Aggregate averages can hide substantial heterogeneity across driving windows. To make this variability explicit, Figure~\ref{fig:highd_errdist} plots the per-window distributions of RMSE and ES for the two ASBC variants. Two observations stand out. First, the error distributions are heavy-tailed---especially for spacing $s$---indicating that a non-trivial subset of windows is genuinely difficult (e.g., abrupt stop-and-go transitions). Second, the Mat\'ern residual model yields a systematic downward shift and a tighter spread in both RMSE and ES. For spacing $s$, the median per-window RMSE decreases from 3.19 to 1.75, and the median ES decreases from 17 to 9.84; similar ES improvements appear for $v$ and $a$. These distribution-level results complement Table~\ref{tab:highd_summary} by showing that the gains are not driven by a small subset of easy windows.

\begin{figure*}[!t]
  \centering
  \includegraphics{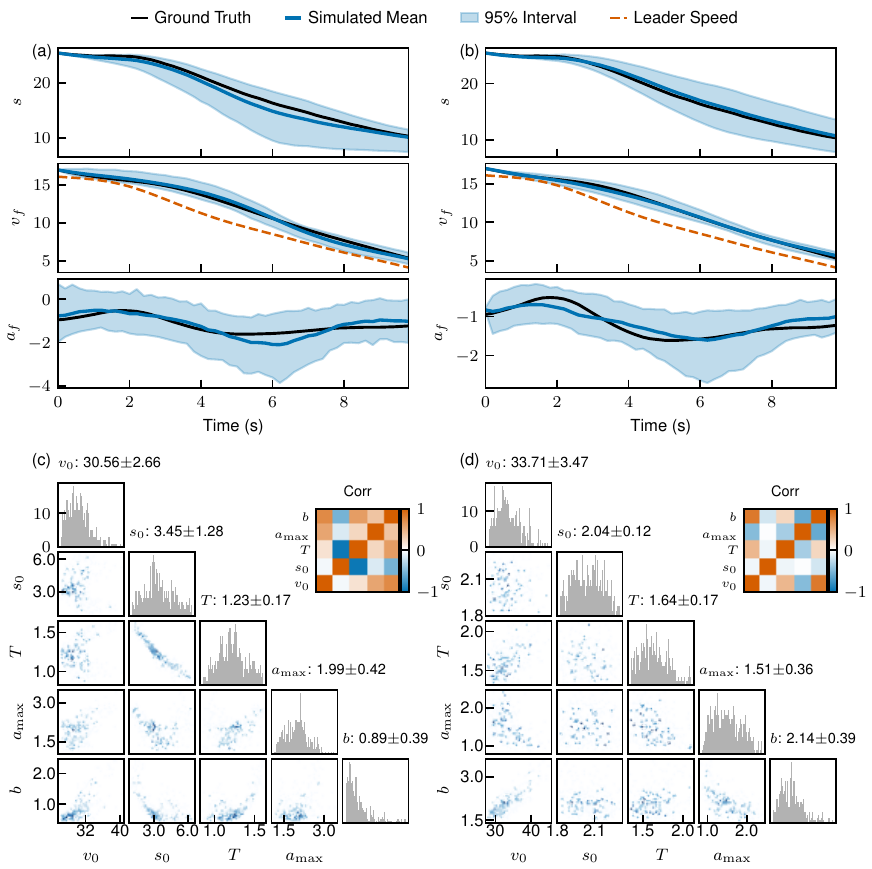}
  \caption{Case study on HighD pair \#426. (a) ASBC-Gaussian and (b) ASBC-Mat\'ern: posterior predictive rollouts with 95\% PI (dashed: aligned leader speed). (c--d) Inferred posterior structure for the same pair, shown via pairwise marginals and the parameter correlation matrix.}
  \label{fig:case_pair426}
\end{figure*}

\subsection{Qualitative case study}
\label{sec:highd_cases}

To complement the aggregate window-level results, we present a representative leader--follower pair (\#426) from the HighD test split. Because HighD does not provide ground-truth driver parameters, this case study is not intended to validate the correctness of inferred parameters. Instead, it serves two purposes: (i) posterior predictive checking against the observed trajectory, and (ii) illustrating how different residual assumptions shift the attribution of mismatch between driver parameters and temporally structured disturbances.

\subsubsection{Posterior predictive rollouts with 95\% prediction intervals}
\label{sec:case_rollouts}

Figure~\ref{fig:case_pair426}(a)--(b) shows posterior predictive rollouts over the same 10\,s evaluation window.
We plot the observed follower trajectories (gap $s$, speed $v_f$, acceleration $a_f$), the posterior predictive mean, and the 95\% prediction interval (PI) computed from posterior samples. The aligned leader speed (orange dashed) is treated as a known exogenous input.

Both variants track the dominant short-horizon trends in the predictive mean, but their uncertainty evolves differently over time. Under ASBC-Gaussian, the PI boundaries fluctuate more pointwise (most clearly in acceleration). In contrast, ASBC-Mat\'ern produces uncertainty that evolves more smoothly and expands coherently around sustained transitions (e.g., prolonged braking), consistent with temporally persistent mismatch rather than independent noise. Such temporal coherence is especially important in stop-and-go regimes, where temporally inconsistent uncertainty can be misleading even when the mean trajectory appears reasonable.

\subsubsection{Posterior structure and attribution under correlated residuals}
\label{sec:case_posterior}
Figure~\ref{fig:case_pair426}(c)--(d) summarizes the inferred posterior structure over the five core IDM parameters.\footnote{For comparability, we focus on the core IDM parameters; both variants also infer residual hyperparameters, which are omitted here.} We interpret these posteriors as explanatory summaries that support the predictive behavior in Fig.~\ref{fig:case_pair426}(a)--(b), rather than as verifiable estimates of true driver traits.

The key mechanism is an attribution trade-off. Because the residual enters through the acceleration update and is integrated into future speed and spacing, persistent deviations over a short segment can be explained either by shifting $\boldsymbol{\theta}$ or by allowing a temporally structured residual path $\{r_t\}$ to absorb systematic mismatch. This choice can also influence the predictive mean. Under ASBC-Mat\'ern, conditioning on an initial residual prefix can induce a nonzero conditional mean for future residuals, which can shift the predictive mean even when $\boldsymbol{\theta}$ is held fixed. Under an i.i.d.\ Gaussian residual, the conditional mean of future noise remains zero, so systematic mean shifts are primarily expressed through $\boldsymbol{\theta}$.

For pair \#426, the Gaussian variant places more mass on a larger jam distance and weaker comfortable braking (Fig.~\ref{fig:case_pair426}c), whereas the Mat\'ern variant concentrates on a tighter $s_0$ and shifts toward stronger braking and larger headway (Fig.~\ref{fig:case_pair426}d). The correlation matrices further reveal compensatory structure typical in IDM calibration, such as trade-offs between headway $T$ and jam distance $s_0$, and between braking $b$ and acceleration capability $a_{\max}$. Overall, this example illustrates that the residual assumption influences not only the temporal structure of predictive uncertainty but also how discrepancy is allocated between driver parameters and temporally structured mismatch.

\subsection{Diagnostics and stability}
\label{sec:diagnostics}

\begin{figure}[!t]
  \centering
  \includegraphics{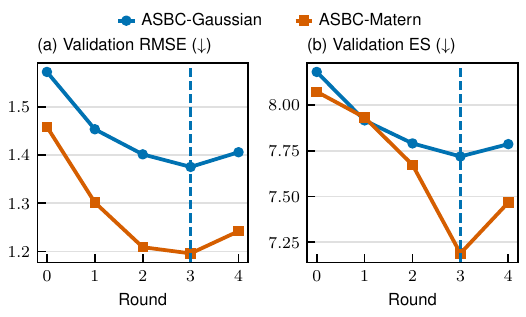}
   \caption{Active-loop convergence on HighD across rounds. The dashed line marks the round selected by the same validation-loss early-stopping rule used in training.}
  \label{fig:highd_convergence}
\end{figure}

\begin{figure}[!t]
  \centering
  \includegraphics{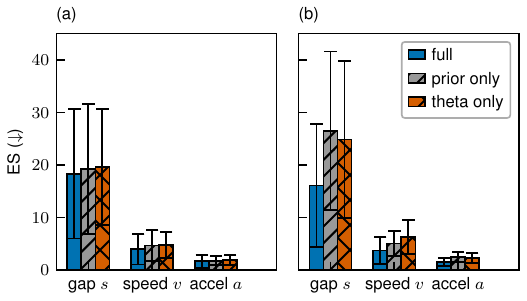}
\caption{Ablations on HighD (ES). We compare the full method against \emph{prior-only} sampling (no soft-shrink) and \emph{$\boldsymbol{\theta}$-only} selection (no joint $(\boldsymbol{\theta},L)$ acquisition). (a) ASBC-Gaussian; (b) ASBC-Mat\'ern.}
  \label{fig:highd_ablation_es}
\end{figure}

\begin{figure}[!t]
  \centering
  \includegraphics[width=0.5\textwidth]{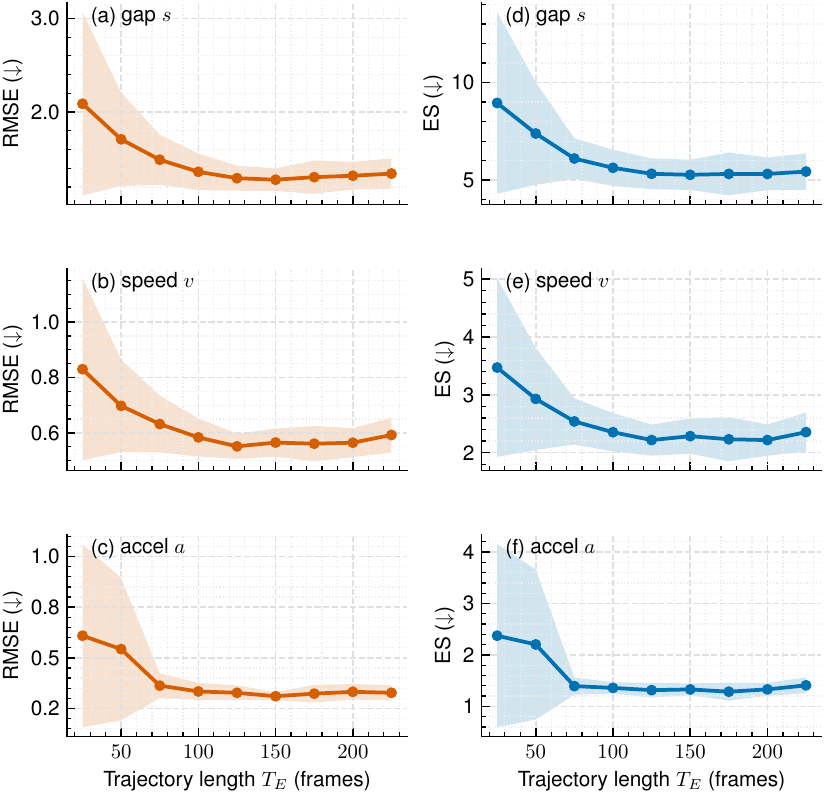}
\caption{Encoder history-length sweep on HighD for ASBC-Gaussian. Mean $\pm$ stand deviation across five seeds.}
  \label{fig:highd_history_sweep}
\end{figure}

This section reports diagnostics on (i) the stability of the active learning loop across acquisition rounds and (ii) sensitivity to key design choices. We present convergence curves with the round selected by early stopping, targeted ablations that isolate acquisition components, and an encoder history-length sweep.

\subsubsection{Active loop convergence curves}
\label{sec:convergence_curves}
Figure~\ref{fig:highd_convergence} shows how evaluation metrics evolve across acquisition rounds for the two residual models. The active loop is monitored and early-stopped using a validation loss on the posterior model; because real trajectories do not provide ground-truth parameters, this loss is computed on a fixed synthetic hold-out set of simulated $(\boldsymbol{\theta},\mathbf{x})$ pairs where $\boldsymbol{\theta}$ is known by construction. To confirm that this stopping rule aligns with the downstream goal, we additionally plot RMSE and ES on held-out real leader--follower pairs. Unless otherwise stated, the reported RMSE/ES are averaged over gap $s$, speed $v$, and acceleration $a$. Encouragingly, the selected round (dashed line) occurs where RMSE and ES have largely plateaued.

A second pattern is the different late-round behavior across residual models. ASBC-Gaussian shows noticeable degradation after the selected round, with both RMSE and ES increasing in the next round. This is plausible in sequential SBI with active acquisition: as the proposed and selected $(\boldsymbol{\theta},L)$ queries become more concentrated, later rounds can shift the effective training distribution away from the evaluation domain, leading to over-specialization or instability. In contrast, ASBC-Mat\'ern reaches its best region early and remains relatively stable thereafter, consistent with a correlated residual that can absorb temporally structured mismatch without pushing the posterior over IDM parameters to drift. Overall, the curves suggest that most gains occur in the first few rounds and that cross-round early stopping is a useful guardrail against diminishing returns and late-round degradation.

\subsubsection{Ablations and sensitivity analyses}
\label{sec:ablations_sensitivity}

\paragraph{Ablating acquisition components}
To identify which components of the active design drive predictive quality, we compare the full method against two ablations (Fig.~\ref{fig:highd_ablation_es}). The \emph{prior-only} variant disables the soft-shrink proposal and samples parameters purely from the prior. The \emph{$\boldsymbol{\theta}$-only} variant removes joint $(\boldsymbol{\theta},L)$ acquisition by pairing selected parameters with random leader windows, thereby dropping scenario representativeness and joint-space diversity control. Both ablations degrade distributional accuracy, with a markedly larger effect under the correlated residual model: relative to each ablation baseline, the full method reduces ES by 26.5--41\% under ASBC-Mat\'ern versus 4.6--17.5\% under ASBC-Gaussian across variables and ablations, where the relative change is computed as $(\mathrm{ES}_{\text{abl}}-\mathrm{ES}_{\text{full}})/\mathrm{ES}_{\text{abl}}\times 100\%$.
This pattern supports a key takeaway: temporally correlated residuals can absorb persistent mismatch, making parameter identification more sensitive to the choice of leader scenarios. Joint acquisition mitigates this by selecting informative leader contexts, which helps disentangle structured residual effects from driver parameters; this benefit is less critical under i.i.d.\ Gaussian noise, where mismatches tend to average out more easily.

\paragraph{Encoder history-length sweep}
We additionally sweep the encoder history length $T_E$ for ASBC-Gaussian to quantify how much past context is needed to stabilize posterior inference (Figure~\ref{fig:highd_history_sweep}). Performance improves sharply as $T_E$ increases from very short histories and then saturates: beyond a moderate $T_E$, extending the encoder context yields limited marginal gains. Notably, seed-to-seed variability is largest at small $T_E$ and shrinks substantially once the encoder observes sufficient history to disambiguate driving regimes. We therefore use the default $T_E=75$ frames (15s) in all main experiments, and treat this sweep as a diagnostic that our configuration operates in a stable regime rather than a finely tuned edge case.

\subsection{Discussion}
\label{sec:discussion}

\paragraph{Key takeaways}
Our experiments suggest three takeaways for uncertainty-aware short-horizon car-following simulation. 
\textbf{(1) Use window-level distributions and proper distributional scores:} averages can mask heavy-tailed behavior in short-horizon simulation, whereas window-level error distributions and proper distributional scores such as ES better reflect fidelity under uncertainty.
\textbf{(2) Correlated residuals improve temporal coherence:} compared with i.i.d.\ Gaussian residuals, the Mat\'ern model yields prediction intervals that evolve more smoothly and expand coherently around sustained transitions.
\textbf{(3) Acquisition and residual structure interact:} joint $(\boldsymbol{\theta},L)$ acquisition is especially beneficial under correlated residuals, consistent with the intuition that persistent mismatch makes parameter identification more sensitive to scenario choice.

A common evaluation in synthetic calibration studies is parameter recovery: whether the inferred posterior recovers a known ground-truth parameter distribution (e.g., ABC-based calibration for automated-vehicle controllers~\cite{jiang2025abcAV}). This is appropriate when the simulator is treated as the data-generating mechanism and parameters are the primary object of interest. Our setting is different. For naturalistic human driving data (e.g., HighD), there is no observable ground-truth driver parameter vector, and multiple combinations of parameters and residual realizations can explain similar short-horizon behavior. Accordingly, we interpret inferred parameters as \emph{predictive} summaries rather than verifiable driver traits, and we evaluate performance through posterior predictive checks and trajectory-level metrics under a fixed leader input.

\paragraph{Implementation guidance}
In practice, we recommend the following workflow:
(i)~use a fixed short-horizon protocol (e.g., 10\,s rolling windows with reset initial states);
(ii)~draw enough posterior samples so ES and prediction-interval estimates are stable;
(iii)~apply cross-round early stopping using a fixed synthetic hold-out set, since real trajectories lack ground-truth $\boldsymbol{\theta}$;
(iv)~maintain a leader-window bank that balances representativeness and diversity (optionally with mild augmentation and strict physical validity checks);
(v)~avoid overly short conditioning windows, which increase seed sensitivity.

\paragraph{Open problems}
Our study mainly exploits \,\emph{amortization} in a residual-augmented IDM setting. A natural next step is to extend to black-box microscopic simulators (e.g., SUMO~\cite{krajzewicz2012recent}), where likelihood-free learning is essential and amortization matters even more for scale.

Two open problems become central in that setting:
\textbf{(1) Objectives under simulator--reality gaps.} Misspecification can dominate calibration results, so it is unclear which discrepancy measures best reflect intended network-level behavior. A key question is how to design robust, task-aligned objectives beyond short-horizon trajectory errors (e.g., distributional targets for travel times, queues, or speed fields) while remaining sensitive to behavioral mechanisms.
\textbf{(2) Scenario design at scale.} The scenario space (demand, signals, geometry, incidents) is much larger than a leader-window stimulus, motivating extensions of joint acquisition from $(\boldsymbol{\theta},L)$ to $(\boldsymbol{\theta},\text{scenario})$. A key direction is to design scenario parameterizations and representativeness/diversity criteria that remain effective under limited simulation budgets.

A separate limitation is fixed-length conditioning. While fixed windows stabilize training and evaluation, longer histories are not always more informative for parameter identification: long free-flow segments or rapidly switching regimes can dilute the interactions that carry parameter information. Promising directions include learning data-adaptive trajectory summaries (e.g., attention pooling or event-centric encoders~\cite{ilse2018attention,lee2019set}) and learning data-dependent window selection $W(\mathbf{x})$ so the model can focus on decisive interactions such as closing and braking events.

\section{Conclusion and Future Work}
\label{sec:conclusion}

This paper presents an active, amortized simulation-based inference framework for calibrating stochastic car-following models from trajectory data. The central idea is to replace per-trajectory optimization or MCMC with a trajectory-conditioned neural posterior estimator: once trained, driver-specific posteriors are produced in a single forward pass, enabling stochastic calibration at dataset scale. To make amortized SBI practical under limited simulation budgets, we introduce a joint $(\boldsymbol{\theta},L)$ acquisition strategy that selects both parameters and representative leader-trajectory windows, concentrating simulation effort where it most improves posterior quality. On HighD, the resulting posterior predictive simulations outperform pooled Bayesian baselines in fidelity and distributional scores, and the Mat\'ern-$5/2$ residual variant yields more temporally coherent uncertainty and better distributional alignment.

The proposed framework provides a scalable route to calibrated uncertainty for microscopic simulation, which matters when safety and reliability hinge on tail events rather than average behavior. Fast driver-specific posterior inference further enables uncertainty-aware rollouts and stress testing across diverse interactions, supporting robust scenario evaluation and simulation-based decision support in traffic engineering. Promising future directions include deployment to black-box microscopic simulators and the development of data-adaptive trajectory representations.

\section*{Acknowledgements}
This research is supported by the Natural Sciences and Engineering Research Council of Canada (NSERC) of Canada. The authors would also like to thank the McGill Engineering Doctoral Awards (MEDA), the Interuniversity Research Centre on Enterprise Networks, Logistics and Transportation (CIRRELT), and Fonds de recherche du Québec -- Nature et technologies (FRQNT) for providing scholarships and funding to support this study.

\bibliographystyle{IEEEtran}
\bibliography{references}

\clearpage
\onecolumn
\appendices
\addcontentsline{toc}{section}{Appendices}

\section*{Appendix Overview}
\noindent This appendix contains the following supplemental materials:
\begin{itemize}
  \item Appendix~\ref{app:encoder_details}: Encoder architecture details (Local--Global Transformer and context embedding).
  \item Appendix~\ref{app:flow_transform}: Posterior flow and parameter transform (MAF density, positive transform, and sampling).
  \item Appendix~\ref{app:npe_objective}: Derivation of the NPE objective (connection to an expected posterior KL).
  \item Appendix~\ref{app:leader_aug}: Leader-window augmentation and filtering (bank construction, augmentations, and plausibility checks).
  \item Appendix~\ref{app:divpen}: Diversity penalty in joint acquisition (distances and kernel penalty).
  \item Appendix~\ref{app:alpha_score}: MC-dropout uncertainty score for ASNPE.
  \item Appendix~\ref{app:proposal}: Posterior-informed proposal construction for the soft-shrink mixture.
  \item Appendix~\ref{app:priors}: Prior specification and feasibility constraints (IDM and residual hyperparameters).
  \item Appendix~\ref{app:impl_hparams}: Implementation hyperparameters and thresholds (training, leader-bank, and acquisition settings).
\end{itemize}
\vspace{0.5em}

\section{Encoder architecture details}
\label{app:encoder_details}

We use a Local--Global Transformer encoder that combines windowed self-attention with a global summary token. The windowed mechanism captures fine-grained local interactions, while a dedicated \textsc{[CLS]} token aggregates information across the full sequence. We incorporate learnable relative positional biases following.

\paragraph*{Input projection}
Given a trajectory $\mathbf{x}=(\mathbf{h}_1,\dots,\mathbf{h}_T)\in\mathbb{R}^{T\times 3}$, we map each input state to a $d$-dimensional representation using a learnable projection
\[
\mathbf{z}_t^{(0)}=\mathbf{W}_{\mathrm{in}}\mathbf{h}_t,\qquad \mathbf{W}_{\mathrm{in}}\in\mathbb{R}^{d\times 3}.
\]

\paragraph*{Local--global attention layers}
We refine representations using a stack of $L$ Local--Global attention layers. At layer $\ell=0,\dots,L-1$, local attention is applied within a fixed-size sliding window $\mathcal{W}(t)$ centered at $t$,
\[
\mathbf{z}_t^{(\ell+1)}=
\textsc{MHA}_{\mathrm{local}}\!\left(
\mathbf{z}_t^{(\ell)},\{\mathbf{z}_{\tau}^{(\ell)}\}_{\tau\in\mathcal{W}(t)},\mathbf{B}
\right).
\]
In parallel, global attention operates on the \textsc{[CLS]} token to aggregate sequence-level information,
\[
\mathbf{z}_{\textsc{[cls]}}^{(\ell+1)}=
\textsc{MHA}_{\mathrm{global}}\!\left(
\mathbf{z}_{\textsc{[cls]}}^{(\ell)},\{\mathbf{z}_{\tau}^{(\ell)}\}_{\tau=1}^{T},\mathbf{B}
\right).
\]

\paragraph*{Context vector}
After $L$ layers, the trajectory embedding is taken as the final hidden state of the \textsc{[CLS]} token,
\[
\mathbf{c}=\mathbf{z}_{\textsc{[cls]}}^{(L)}\in\mathbb{R}^{d}.
\]

\section{Details of the posterior flow and parameter transform}
\label{app:flow_transform}

\subsection{Conditional masked autoregressive flow (MAF)}
\label{app:maf_details}

We model the posterior in an unconstrained space with a conditional masked autoregressive flow (MAF)~\cite{papamakarios2017masked}. Let $\mathbf{u}\in\mathbb{R}^{D}$ denote the unconstrained latent parameter vector and $\mathbf{c}$ the context embedding. The flow defines an invertible mapping between $\mathbf{u}$ and a base variable $\mathbf{z}\in\mathbb{R}^{D}$:
\begin{equation}
\mathbf{z}=f_{\phi}(\mathbf{u};\mathbf{c}), \qquad \mathbf{z}\sim \mathcal{N}(\mathbf{0},\mathbf{I}),
\end{equation}
where $f_{\phi}(\cdot;\mathbf{c})$ is a composition of $K$ invertible, context-conditioned transformations. Specifically, define $\mathbf{z}_0=\mathbf{u}$ and for $k=1,\dots,K$,
\begin{equation}
\mathbf{z}_k=f_{k,\phi}(\mathbf{z}_{k-1};\mathbf{c}),\qquad \mathbf{z}_K=\mathbf{z}.
\end{equation}
By the change-of-variables formula, the conditional density of $\mathbf{u}$ is
\begin{equation}
q_{\phi}(\mathbf{u}\mid\mathbf{c})
=
\mathcal{N}\!\left(\mathbf{z}_K;\mathbf{0},\mathbf{I}\right)
\prod_{k=1}^{K}
\left|
\det\!\left(\frac{\partial \mathbf{z}_k}{\partial \mathbf{z}_{k-1}}\right)
\right|,
\label{eq:maf_u_density_app}
\end{equation}
and hence
\begin{equation}
\log q_{\phi}(\mathbf{u}\mid\mathbf{c})
=
\log \mathcal{N}\!\left(\mathbf{z}_K;\mathbf{0},\mathbf{I}\right)
+
\sum_{k=1}^{K}
\log\left|
\det\!\left(\frac{\partial \mathbf{z}_k}{\partial \mathbf{z}_{k-1}}\right)
\right|.
\label{eq:maf_u_logdensity_app}
\end{equation}
This expression is used for exact log-density evaluation during training and for the acquisition score in \S\ref{sec:asnpe}. At test time, sampling proceeds by drawing $\mathbf{z}\sim\mathcal{N}(\mathbf{0},\mathbf{I})$ and applying the inverse mapping $\mathbf{u}=f_{\phi}^{-1}(\mathbf{z};\mathbf{c})$.

\subsection{Positive parameter transform}
\label{app:param_transform}

To enforce positivity of the physical parameters, we apply an element-wise bijection
$T:\mathbb{R}^{D}\to\mathbb{R}_{+}^{D}$ defined by
\begin{equation}
\boldsymbol{\theta} = T(\mathbf{u}) = \mathrm{softplus}(\mathbf{u}) + \varepsilon,
\qquad \varepsilon>0,
\end{equation}
where $\mathrm{softplus}(x)=\log(1+\exp(x))$ is applied component-wise. The offset $\varepsilon$ provides a strict lower bound $\theta_i>\varepsilon$ for numerical stability. The inverse is also element-wise:
\begin{equation}
u_i = T^{-1}(\theta_i) = \log\!\left(\exp(\theta_i-\varepsilon)-1\right),
\qquad i=1,\dots,D.
\end{equation}

The Jacobian matrix $\partial \boldsymbol{\theta}/\partial \mathbf{u}$ is diagonal with
\begin{equation}
\frac{\partial \theta_i}{\partial u_i} = \sigma(u_i),
\qquad \sigma(u)=\frac{1}{1+\exp(-u)},
\end{equation}
hence
\begin{equation}
\log\left|\det\left(\frac{\partial \boldsymbol{\theta}}{\partial \mathbf{u}}\right)\right|
= \sum_{i=1}^{D}\log \sigma(u_i).
\end{equation}

Combining the flow density in \eqref{eq:maf_u_density_app} with the transform $T(\cdot)$, the physical-space posterior density is
{\small
\begin{equation}
\log q_{\phi}(\boldsymbol{\theta}\mid \mathbf{c})
=
\log q_{\phi}(\mathbf{u}\mid \mathbf{c})
-
\log\left|\det\left(\frac{\partial \boldsymbol{\theta}}{\partial \mathbf{u}}\right)\right|,
\qquad \mathbf{u}=T^{-1}(\boldsymbol{\theta}).
\label{eq:theta_density_app}
\end{equation}
}

\subsection{Posterior sampling in the physical space}
\label{app:sampling}

Given a context $\mathbf{c}$, posterior samples are obtained by
{\footnotesize
\begin{equation}
\begin{aligned}
\mathbf{z}^{(s)} &\sim \mathcal{N}(\mathbf{0},\mathbf{I}),\\
\mathbf{u}^{(s)} &= f_{\phi}^{-1}(\mathbf{z}^{(s)};\mathbf{c}),\\
\boldsymbol{\theta}^{(s)} &= T(\mathbf{u}^{(s)}),\qquad s=1,\dots,S.
\end{aligned}
\end{equation}
}
The same procedure applies under both residual specifications; only the dimensionality $D$ of $\boldsymbol{\theta}$ differs (see \S\ref{sec:prior}).

\section{Derivation of the NPE objective}
\label{app:npe_objective}

This appendix provides a short derivation connecting the maximum-likelihood objective used in neural posterior estimation to a Kullback--Leibler (KL) divergence between the true posterior and the learned approximation. We present the derivation at the level of the joint distribution induced by simulation and encoding, which is the setting used in \S\ref{sec:objective}.

\subsection{From conditional likelihood to posterior KL}

Let $\mathbf{x}$ denote a simulated follower trajectory and let $\mathbf{c}=f_{\psi}(\mathbf{x})$ be its context embedding. Consider any joint distribution over simulated pairs $(\boldsymbol{\boldsymbol{\theta}},\mathbf{c})$ with density
\begin{equation}
p(\boldsymbol{\boldsymbol{\theta}},\mathbf{c}) = p(\mathbf{c})\,p(\boldsymbol{\boldsymbol{\theta}}\mid \mathbf{c}).
\label{eq:app_joint_factor}
\end{equation}
Given a conditional density model $q_{\phi}(\boldsymbol{\boldsymbol{\theta}}\mid \mathbf{c})$, the standard NPE objective maximizes the expected conditional log-likelihood
\begin{equation}
\mathcal{L}(\psi,\phi)
=
\mathbb{E}_{(\boldsymbol{\boldsymbol{\theta}},\mathbf{c})\sim p}
\left[
\log q_{\phi}(\boldsymbol{\boldsymbol{\theta}}\mid \mathbf{c})
\right].
\label{eq:app_npe_obj}
\end{equation}
Using the factorization in \eqref{eq:app_joint_factor}, we can rewrite
{\footnotesize
\begin{align}
\mathcal{L}(\psi,\phi)
&=
\mathbb{E}_{\mathbf{c}\sim p(\mathbf{c})}
\left[
\mathbb{E}_{\boldsymbol{\boldsymbol{\theta}}\sim p(\boldsymbol{\boldsymbol{\theta}}\mid \mathbf{c})}
\log q_{\phi}(\boldsymbol{\boldsymbol{\theta}}\mid \mathbf{c})
\right] \nonumber\\
&=
\mathbb{E}_{\mathbf{c}\sim p(\mathbf{c})}
\left[
\mathbb{E}_{\boldsymbol{\boldsymbol{\theta}}\sim p(\boldsymbol{\boldsymbol{\theta}}\mid \mathbf{c})}
\log p(\boldsymbol{\boldsymbol{\theta}}\mid \mathbf{c})
\right]
-
\mathbb{E}_{\mathbf{c}\sim p(\mathbf{c})}
\mathrm{KL}\!\left(
p(\boldsymbol{\boldsymbol{\theta}}\mid \mathbf{c}) \,\|\, q_{\phi}(\boldsymbol{\boldsymbol{\theta}}\mid \mathbf{c})
\right).
\label{eq:app_kl_decomp}
\end{align}}
The first term in \eqref{eq:app_kl_decomp} does not depend on $\phi$ and is therefore a constant with respect to the posterior model. Hence, maximizing $\mathcal{L}$ is equivalent to minimizing an expected KL divergence between the true conditional $p(\boldsymbol{\theta}\mid \mathbf{c})$ and the learned approximation $q_{\phi}(\boldsymbol{\theta}\mid \mathbf{c})$, averaged over $\mathbf{c}\sim p(\mathbf{c})$.

\subsection{Interpretation under sequential acquisition}

In our sequential training procedure, the training pairs $(\boldsymbol{\theta},\mathbf{c})$ are not sampled from a fixed prior alone. Instead, at each round $r$ we generate a dataset of pairs by selecting parameter--scenario candidates, simulating trajectories under the residual-augmented calibration model, and encoding them into contexts. This induces an empirical distribution $\mathcal{D}_r$ over $(\boldsymbol{\theta},\mathbf{c})$, and we train by maximizing
\[
\mathcal{L}(\psi,\phi)=
\mathbb{E}_{(\boldsymbol{\theta},\mathbf{c})\sim \mathcal{D}_r}
\left[
\log q_{\phi}(\boldsymbol{\theta}\mid \mathbf{c})
\right],
\]
as stated in \eqref{eq:npe_loss_final}. From \eqref{eq:app_kl_decomp}, this corresponds to fitting $q_{\phi}(\boldsymbol{\theta}\mid \mathbf{c})$ to the conditional distribution implied by the current round's acquisition and simulation process. In practice, this is appropriate because the active module is designed to concentrate simulation effort on parameter--scenario pairs that are informative for learning the posterior mapping on the observed domain of interest. We implement this sequential refinement using a fixed-capacity buffer that mixes newly acquired pairs with recent history to stabilize training.

\section{Leader-window augmentation and filtering}
\label{app:leader_aug}

This appendix summarizes how we construct the leader-window bank used for joint acquisition. The goal is to provide a diverse set of realistic driving contexts while keeping the simulation inputs physically plausible and consistent with the sampling rate $\Delta t$.

\subsection{Leader windows and bank construction}
From each observed leader trajectory $\boldsymbol{\ell}^{(n)}$, we extract overlapping windows of a fixed length $W$ with stride $S$. Each window is denoted by $L$ and contains a leader speed profile $\{v_{\ell}(t)\}_{t=1}^{W}$; when acceleration is not directly available, we compute it by finite differences,
\[
a_{\ell}(t) = \frac{v_{\ell}(t)-v_{\ell}(t-1)}{\Delta t},
\qquad t=2,\ldots,W,
\]
and set $a_{\ell}(1)=a_{\ell}(2)$ for convenience. The resulting set of real windows defines $\mathcal{L}_{\mathrm{real}}$.

To increase context diversity without introducing unrealistic motion, we generate an augmented set $\mathcal{L}_{\mathrm{syn}}$ by applying lightweight transformations to windows from $\mathcal{L}_{\mathrm{real}}$, followed by deterministic plausibility checks. The round-dependent leader sampling law is then
\[
\mathcal{L}_r = (1-\alpha_r)\,\mathcal{L}_{\mathrm{real}} + \alpha_r\,\mathcal{L}_{\mathrm{syn}},
\]
as stated in \eqref{eq:leader_mix_new}.

\subsection{Physics-respecting augmentations}
We apply a small set of augmentations designed to preserve the overall shape of leader motion while introducing mild variability.

\paragraph{A1) Band-limited speed perturbation}
We add a low-amplitude perturbation $\delta v(t)$ to the leader speed and obtain
\[
v'_{\ell}(t) = v_{\ell}(t) + \delta v(t),
\]
where $\delta v(t)$ is generated by filtering white noise to retain only low-to-mid frequency components. This avoids injecting unrealistic high-frequency jitter while allowing slightly different stop-and-go patterns.

\paragraph{A2) Global rescaling}
We apply a mild global scaling to speed and the implied acceleration,
\[
v'_{\ell}(t)=\kappa\,v_{\ell}(t), \qquad \kappa \in [1-\beta_{\max},\,1+\beta_{\max}],
\]
with $\beta_{\max}$ set to a small value. This preserves temporal structure while adjusting the overall intensity of the leader motion.

\paragraph{A3) Piecewise-affine time warping}
To introduce small timing variations, we optionally apply a piecewise-affine reparameterization of time. Let $\tau(t)$ be a monotone mapping from $\{1,\ldots,W\}$ to $\{1,\ldots,W\}$ constructed by stretching or compressing a few subsegments by a small factor and then linearly interpolating. The warped profile is
\[
v'_{\ell}(t) = v_{\ell}\!\big(\tau(t)\big),
\]
implemented with linear interpolation on the discrete grid.

In all cases, we recompute $a'_{\ell}(t)$ from $v'_{\ell}(t)$ using the same finite-difference rule.

\subsection{Filtering and plausibility checks}
All candidate augmented windows are filtered by deterministic constraints to ensure physical plausibility and consistency with the sampling rate:
\begin{itemize}
  \item \textbf{Non-negativity:} $v'_{\ell}(t)\ge 0$ for all $t$.
  \item \textbf{Acceleration bounds:} $|a'_{\ell}(t)| \le a_{\max}^{\mathrm{phys}}$ for all $t$.
  \item \textbf{Jerk bounds:} $\left| \frac{a'_{\ell}(t)-a'_{\ell}(t-1)}{\Delta t} \right| \le j_{\max}$ for all $t\ge 2$.
  \item \textbf{Kinematic consistency:} $\left| v'_{\ell}(t)-v'_{\ell}(t-1)-a'_{\ell}(t)\Delta t \right| \le \varepsilon_{\mathrm{kin}}$ for all $t\ge 2$.
  \item \textbf{Envelope constraints:} $(v'_{\ell}(t),a'_{\ell}(t))$ stays within empirical percentile ranges estimated from the dataset, to avoid rare artifacts.
\end{itemize}
Only windows passing all checks are included in $\mathcal{L}_{\mathrm{syn}}$.

\section{Diversity penalty in joint acquisition}
\label{app:divpen}

This appendix details the diversity penalty $\mathrm{DivPen}(\cdot)$ used in the joint acquisition score in \eqref{eq:joint_score_new}. The purpose is to discourage selecting near-duplicate parameter--scenario pairs within a round, thereby improving coverage of both the parameter space and the leader-context space.

\subsection{Leader feature distance}
Each leader window $L$ is mapped to a fixed feature vector $\varphi(L)\in\mathbb{R}^{d_\ell}$. In our implementation, $\varphi(L)$ includes summary statistics of the leader speed and acceleration over the window, such as means, standard deviations, and extrema. Distances between leader windows are computed using the Euclidean metric in feature space:
\[
d_L(L,L') = \|\varphi(L)-\varphi(L')\|_2.
\]

\subsection{Parameter-space distance}
To compare parameter vectors with different units and scales, we rescale distances in the parameter space using a diagonal or full covariance proxy $\Sigma_{\boldsymbol{\theta}}$. Let $\Sigma_{\boldsymbol{\theta}}^{-1/2}$ be a whitening transform. The rescaled distance is
\[
d_{\boldsymbol{\theta}}(\boldsymbol{\theta},\boldsymbol{\theta}') =
\|\Sigma_{\boldsymbol{\theta}}^{-1/2}(\boldsymbol{\theta}-\boldsymbol{\theta}')\|_2.
\]
In practice, $\Sigma_{\boldsymbol{\theta}}$ can be set using prior scales or empirical standard deviations of recent samples.

\subsection{Kernel dispersion penalty}
Given a current selected set $S$ of pairs in a greedy round, we penalize selecting a new candidate $(\boldsymbol{\theta},L)$ that is too close to the existing set. We implement
{\footnotesize
\begin{equation}
\label{eq:divpen_kernel}
\mathrm{DivPen}\!\big(S\cup\{(\boldsymbol{\theta},L)\}\big)
=
\frac{1}{|S|}\sum_{(\boldsymbol{\theta}',L')\in S}
\exp\!\left(
-\frac{d_L(L,L')^2}{\tau_L^2}
-\frac{d_{\boldsymbol{\theta}}(\boldsymbol{\theta},\boldsymbol{\theta}')^2}{\tau_{\boldsymbol{\theta}}^2}
\right),
\end{equation}}
where $\tau_L$ and $\tau_{\boldsymbol{\theta}}$ are temperature parameters controlling how quickly similarity decays in leader space and parameter space, respectively. When $S=\varnothing$, we set $\mathrm{DivPen}(\{(\boldsymbol{\theta},L)\})=0$.

\subsection{Greedy selection with diversity}
Starting from $S=\varnothing$, we iteratively add the candidate pair that maximizes the joint acquisition score $s_r(\boldsymbol{\theta},L)$ in \eqref{eq:joint_score_new}. The penalty in \eqref{eq:divpen_kernel} encourages the selected set to spread across distinct leader contexts and distinct parameter regions, which improves coverage and reduces redundant simulations within a fixed budget.

\section{MC-dropout uncertainty score for ASNPE}
\label{app:alpha_score}

This appendix specifies the MC-dropout score $\alpha_r(\boldsymbol{\theta})$ used in the joint acquisition step. Let $\mathbf{c}$ denote a context embedding computed from an observed trajectory, and let $\{\phi_r^{(m)}\}_{m=1}^{M}$ be $M$ dropout realizations of the posterior network at round $r$. For any candidate parameter $\boldsymbol{\theta}$, we evaluate the log-density under each dropout realization,
\[
\ell_r^{(m)}(\boldsymbol{\theta};\mathbf{c})=\log q_{\phi_r^{(m)}}(\boldsymbol{\theta}\mid \mathbf{c}).
\]
We compute the mean and variance across dropout samples.
{\small
\[
\mu_r(\boldsymbol{\theta};\mathbf{c})=\frac{1}{M}\sum_{m=1}^{M}\ell_r^{(m)}(\boldsymbol{\theta};\mathbf{c}),
\qquad
v_r(\boldsymbol{\theta};\mathbf{c})=\mathrm{Var}_{m}\!\left[\ell_r^{(m)}(\boldsymbol{\theta};\mathbf{c})\right].
\]}
We then define an epistemic-uncertainty score as
\begin{equation}
\alpha_r(\boldsymbol{\theta};\mathbf{c}) = \mu_r(\boldsymbol{\theta};\mathbf{c}) + \log\!\big(v_r(\boldsymbol{\theta};\mathbf{c})+\varepsilon_{\alpha}\big),
\label{eq:alpha_mc_dropout_app}
\end{equation}
where $\varepsilon_{\alpha}>0$ is a small constant for numerical stability.

In practice, we compute \eqref{eq:alpha_mc_dropout_app} on a subset of context embeddings obtained from a subset of observed trajectories $\mathcal{X}^{\mathrm{sub}}\subset \mathcal{X}$ and aggregate the scores across that subset. Specifically, for each $\mathbf{x}\in\mathcal{X}^{\mathrm{sub}}$ we compute $\mathbf{c}=f_{\psi}(\mathbf{x})$ and take
\[
\alpha_r(\boldsymbol{\theta})=\frac{1}{|\mathcal{X}^{\mathrm{sub}}|}\sum_{\mathbf{x}\in\mathcal{X}^{\mathrm{sub}}}\alpha_r\!\big(\boldsymbol{\theta}; f_{\psi}(\mathbf{x})\big).
\]
For numerical robustness, the resulting $\alpha_r(\boldsymbol{\theta})$ is standardized across candidate parameters before ranking.

\section{Posterior-informed proposal construction}
\label{app:proposal}

This appendix describes how we construct the posterior-informed proposal $\tilde q_{r-1}(\boldsymbol{\theta})$ used in the soft-shrink mixture \eqref{eq:soft_proposal_new}. At round $r{-}1$, we select a subset of observed trajectories $\mathcal{X}^{\mathrm{sub}} \subset \mathcal{X}$, either as a fixed validation subset or as a randomly sampled subset with a fixed seed for reproducibility. For each trajectory $\mathbf{x}\in \mathcal{X}^{\mathrm{sub}}$, we draw posterior samples $\boldsymbol{\theta}^{(s)} \sim q_{\phi_{r-1}}(\boldsymbol{\theta}\mid \mathbf{x})$ and pool them to form an empirical proposal distribution. Conceptually, this corresponds to mixing the per-trajectory posteriors over $\mathcal{X}^{\mathrm{sub}}$,
\[
\tilde q_{r-1}(\boldsymbol{\theta})\approx \frac{1}{|\mathcal{X}^{\mathrm{sub}}|}\sum_{\mathbf{x}\in\mathcal{X}^{\mathrm{sub}}} q_{\phi_{r-1}}(\boldsymbol{\theta}\mid \mathbf{x}),
\]
which we implement by maintaining the pooled sample set and resampling from it when generating parameter candidates.

\section{Prior specification and feasibility constraints}
\label{app:priors}

\subsection{IDM parameter priors}
\label{app:priors_idm}

We place priors in log space to ensure positivity. 
For $\boldsymbol{\theta}_{\mathrm{IDM}}=[v_0,s_0,T,a_{\max},b]^\top$, we use
\begin{equation}
\log \boldsymbol{\theta}_{\mathrm{IDM}} \sim 
\mathcal{N}\!\left(\log \boldsymbol{\theta}_{\mathrm{rec}},\; \mathrm{diag}(\sigma_{\log}^2)\right),
\label{eq:prior_idm_app}
\end{equation}
with $\boldsymbol{\theta}_{\mathrm{rec}}=[33.3,\,2.0,\,1.6,\,1.5,\,1.67]^\top$~\cite{treiber2012trajectory} and $\sigma_{\log}=\mathbf{1}$.
To promote physically plausible and numerically stable simulations, we truncate the prior support to feasible ranges (Table~\ref{tab:prior_ranges}) and enforce these constraints via rejection sampling at simulation time.

\begin{table}[t]
\centering
\caption{Feasible ranges used to truncate the IDM prior (enforced during simulation by rejection sampling).}
\label{tab:prior_ranges}
\begin{tabular}{lc}
\toprule
Parameter & Range \\
\midrule
$v_0$ (m/s) & $[20,40]$ \\
$s_0$ (m) & $[1,6]$ \\
$T$ (s) & $[0.6,4.5]$ \\
$a_{\max}$ (m/s$^2$) & $[0.2,3.5]$ \\
$b$ (m/s$^2$) & $[0.4,4.0]$ \\
\bottomrule
\end{tabular}
\end{table}

\subsection{Residual hyperparameter priors}
\label{app:priors_residual}

For the i.i.d.\ Gaussian residual model, we place a log-normal prior on the residual acceleration scale,
\begin{equation}
\log \sigma \sim \mathcal{N}(-1.0,\;0.3^2).
\label{eq:prior_sigma_app}
\end{equation}
For the Mat\'ern-$5/2$ residual model, we additionally place a log-normal prior on the correlation length scale $\ell$ measured in seconds,
\begin{equation}
\log \ell \sim \mathcal{N}(\log 3.0,\;0.5^2),
\label{eq:prior_ell_app}
\end{equation}
and use a log-normal prior for the residual scale under the correlated specification,
\begin{equation}
\log \sigma \sim \mathcal{N}(\log 0.3,\;0.5^2).
\label{eq:prior_sigma_matern_app}
\end{equation}

The inference model in \S\ref{sec:npe} conditions on the observed trajectory $\mathbf{x}$ and outputs posterior samples of $\boldsymbol{\theta}$ under either residual specification.

\section{Implementation hyperparameters and thresholds}
\label{app:impl_hparams}

This appendix lists the concrete hyperparameter values and filtering thresholds used in our implementation. Table~\ref{tab:impl_train_eval} summarizes the training budget, active-loop configuration, and evaluation defaults.
The active loop follows Algorithm~\ref{alg:active_amortized_loop} in the main text. Table~\ref{tab:impl_encoder} summarizes the model configurations of the lightweight Transformer-style encoder with windowed local attention.  Table ~\ref{tab:impl_leader_bank} and ~\ref{tab:impl_acq} summarize the thresholds used for leader-bank construction, filtering, and joint acquisition.

\begin{table*}[!t]
\centering
\caption{Training, active loop, and evaluation configuration used in our HighD experiments.}
\label{tab:impl_train_eval}
\begin{tabular}{ll}
\toprule
\textbf{Component} & \textbf{Value} \\
\midrule
Sampling interval & $\Delta t = 0.2~\mathrm{s}$ (5~Hz) \\
Round budget & up to $R=10$ rounds \\
Round-0 warm-up & \texttt{samples\_initial}$=4000$ simulations \\
Per-round candidates & \texttt{samples\_per\_round}$=5000$ \\
Selected pairs per round & \texttt{B}$=2000$ \\
MC-dropout passes & \texttt{M}$=20$ \\
Replay buffer capacity & \texttt{train\_buffer\_size}=\texttt{samples\_initial}$=4000$ (FIFO) \\
Within-round epochs (max) & \texttt{epochs}$=100$ (early stopping within each round) \\
Batch size & \texttt{batch\_size}$=128$ \\
Optimizer & Adam, learning rate \texttt{lr}$=10^{-3}$ \\
Cross-round early stop & \texttt{min\_rounds}$=3$, \texttt{patience\_round}$=1$, \texttt{min\_delta\_round}$=10^{-3}$ \\
Soft-shrink schedule & \texttt{lambda\_schedule\_base} $=[1.0,0.9,\dots,0.1,0]$ \\
Leader-mix schedule & \texttt{alpha\_schedule\_base} $=[0.00,0.10,0.20,0.30,\dots]$ \\
Evaluation posterior samples & \texttt{eval\_n\_samples}$=500$ \\
Eval window protocol & $H=50$ steps (10~s), stride $S=20$ steps (4~s), up to $m=20$ windows/pair \\
Encoder input length & \texttt{encoder\_target\_len}$=75$ steps (15~s) \\
Validation subset size & \texttt{eval\_val\_size}$=200$ \\
\bottomrule
\end{tabular}
\end{table*}

\begin{table}[!t]
\centering
\caption{Encoder hyperparameters.}
\label{tab:impl_encoder}
\begin{tabular}{ll}
\toprule
\textbf{Item} & \textbf{Value} \\
\midrule
Input dimension & 3 \\
Model dimension & $d=64$ \\
Transformer layers & $L=2$ \\
Attention heads & 4 \\
Local window size & \texttt{local\_window}$=4$ \\
Dropout & 0.2 \\
FFN expansion & $4d$ with GELU \\
Pooling & \textsc{[CLS]} token \\
\bottomrule
\end{tabular}
\end{table}

\begin{table}[!t]
\centering
\caption{Leader-bank construction and filtering hyperparameters.}
\label{tab:impl_leader_bank}
\begin{tabular}{ll}
\toprule
\textbf{Item} & \textbf{Value} \\
\midrule
Window length & $W=75$ steps \\
Window stride & 50 steps \\
Synthetic cap & \texttt{L\_syn\_cap}$=10000$ \\
Augmentation parameters & \texttt{time\_scale\_range}$=[0.9,1.1]$, \texttt{vel\_jitter}$=0.20$ \\
 & \texttt{acc\_clip}$=2.5$, \texttt{jerk\_clip}$=5.0$ \\
Physics filters & $|a_\ell|\le a^{\mathrm{phys}}_{\max}=10.0$, $|j_\ell|\le j_{\max}=20.0$ \\
 & kinematic tolerance $\varepsilon_{\mathrm{kin}}=0.5$ \\
Envelope percentiles & $(v_\ell,a_\ell)$ within $(1,99)$ percentiles \\
\bottomrule
\end{tabular}
\end{table}

\begin{table}[!t]
\centering
\caption{Joint acquisition and diversity-penalty settings.}
\label{tab:impl_acq}
\begin{tabular}{ll}
\toprule
\textbf{Item} & \textbf{Value} \\
\midrule
Parameter candidates cap & \texttt{K\_candidates}$=5000$ \\
Pairs per parameter & \texttt{pairs\_per\_theta}$=10$ \\
Representativeness KNN & \texttt{K\_rho}$=5$, $\varepsilon_{\rho}=10^{-3}$ \\
Diversity penalty & $\gamma=0.1$, $\tau_L=1.0$, $\tau_{\boldsymbol{\theta}}=1.0$ \\
\bottomrule
\end{tabular}
\end{table}

\end{document}